# A batch production scheduling problem in a reconfigurable hybrid manufacturing-remanufacturing system

Behdin Vahedi-Nouri[a], Mohammad Rohaninejad[b], Zdeněk Hanzálek [b,*], Mehdi Foumani[c]

[a] *School of Industrial Engineering, College of Engineering, University of Tehran, Tehran, Iran*
[b] *Czech Institute of Informatics, Robotics, and Cybernetics, Czech Technical University in Prague, Prague, Czech Republic*
[c] *School of Intelligent Finance and Business, Xi'an Jiaotong-Liverpool University, Suzhou, China*

**Abstract**

In recent years, remanufacturing of End-of-Life (EOL) products has been adopted by manufacturing sectors as a competent practice to enhance their sustainability and market share. Due to the mass customization of products and high volatility of market, processing of new products and remanufacturing of EOLs in the same shared facility, namely Hybrid Manufacturing-Remanufacturing System (HMRS), is a mean to keep such production efficient. Accordingly, customized production capabilities are required to increase flexibility, which can be effectively provided under the Reconfigurable Manufacturing System (RMS) paradigm. Despite the advantages of utilizing RMS technologies in HMRSs, production management of such systems suffers excessive complexity. Hence, this study concentrates on the production scheduling of an HMRS consisting of non-identical parallel reconfigurable machines where the orders can be grouped into batches. In this regard, Mixed-Integer Linear Programming (MILP) and Constraint Programming (CP) models are devised to formulate the problem. Furthermore, a computationally efficient solution method is developed based on a Logic-based Benders Decomposition (LBBD) approach. The warm start technique is also implemented by providing a decent initial solution to the MILP model. Computational experiments attest to the LBBD method's superiority over the MILP, CP, and warm-started MILP models by obtaining an average gap of about 2%, besides it yields actionable managerial insights for scheduling in HMRSs.

[*] Corresponding author.

## 1. Introduction

In recent decades, the staggering global population growth, globalization, and digitalization have led to a tremendous demand surge in manufactured products. If the current consumption rate under the conventional linear manufacturing paradigm (i.e., produce, consume, dispose) is maintained, 2.3 times the Earth's resources are required by 2050 to meet the demand [1]. As a result, policymakers, manufacturers, and academia have strictly pursued closing material loops under the concept of circular economy [2, 3]. Remanufacturing has been recognized as an effective approach in the circular economy to achieve sustainability [4] and enhance resiliency against unprecedented disruptions in the supply chain like the COVID-19 pandemic [5]. It is defined as comprehensive processes by which an end-of-life (EOL) product is brought to the like-new state with quality and performance equal to or better than a new product [6, 7].

In comparison with new products manufacturing, remanufacturing enjoys cost savings (up to 50%), material savings (up to 70%), energy savings (up to 60%), $CO_2$ emission reduction (up to 87%), and lower prices (up to 40%) with a decent profit margin (around 20%) [8, 9]. Nevertheless, remanufacturing is an emerging market far from reaching its full potential. Based on a study performed by the European Remanufacturing Network (ERN), the annual value of remanufactured products in Europe was estimated at €30 billion in 2016, possible to reach €100 billion in 2030 [10]. Accordingly, different manufacturing sectors like aerospace, automotive, electrical, medical device, machinery, and furniture have adopted remanufacturing to enhance their market share and environmentally responsible practices [11, 12].

Manufacturers have practiced two approaches regarding the implementation of remanufacturing: 1) performing manufacturing of new products from raw materials and remanufacturing of EOL products in separated dedicated facilities; or 2) performing the both utilizing the same shared facility, namely the hybrid manufacturing-remanufacturing system (HMRS) [13]. The former approach is effective if a large quantity of EOL products is ensured. However, when the quantity of EOL products is limited, or their return timing is unknown (e.g., as in make-to-order environments), it leads to significant underutilization. In these situations, HMRSs provide the most promising alternative [13, 14]. Processing both raw materials and EOL parts in an HMRS may necessitate different production capabilities. In this regard, Reconfigurable Manufacturing Systems (RMSs) best fit HMRSs [15-17] since RMSs can quickly and cost-effectively adjust their structure, hardware, and software components to provide customized production capabilities [18, 19]. In RMSs, various configurations with different production capabilities can be attained by adding, removing, or modifying the modules of machines. For instance, Fig. 1(a) shows a robotic additive-subtractive machine located at Czech Technical University in Prague [20] that its arm's head (Fig. 1(b)) can be changed automatically. As a result, different additive and subtractive capabilities can be implemented.

Despite the advantages of utilizing RMS technologies in HMRSs, production management of such systems encounters higher complexity than conventional systems due to the additional reconfiguration decisions. In this regard, this study concentrates on the production scheduling of an HMRS consisting of non-identical parallel reconfigurable machines (stations or production cells). Moreover, manufacturing and/or remanufacturing orders can be grouped into capacity-constrained batches to save machine setup times. Hence, the serial-batch processing of orders is incorporated, where the processing time of a batch is the total processing time of its assigned orders. Accordingly, the decisions regarding (i) the assignment of the orders to the eligible machine-configuration combinations based on the manufacturing and remanufacturing requirements, (ii) batching the assigned orders to each machine, and (iii) the scheduling of the batches on the machines should be simultaneously made to minimize the makespan.

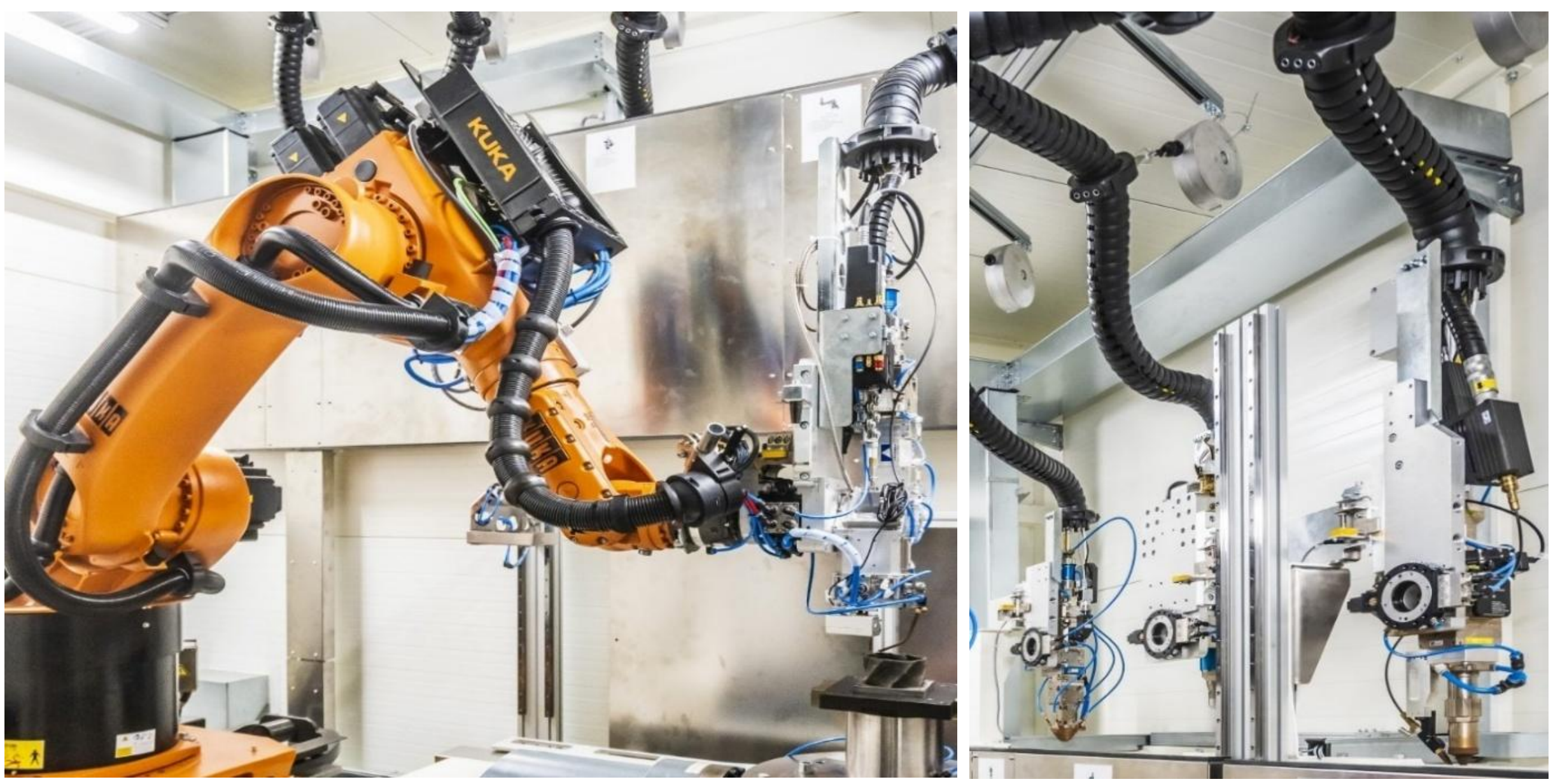

Fig. 1(a). Robot arm

Fig. 1(b). Different heads

**Fig. 1.** A hybrid robotic additive-subtractive machine.

First, the problem is formulated as a Mixed-Integer Linear Programming (MILP) model and its Constraint Programming (CP) counterpart. Furthermore, an efficient solution method is developed based on a Logic-based Benders Decomposition (LBBD) approach. Additionally, to enhance the performance of the MILP model, the warm start technique is implemented by providing an initial solution using a two-step procedure consisting of mathematical models. Finally, through computational experiments, the performance of the developed MILP, CP, LBBD, and warm-started MILP (WMLIP) are compared, and key managerial insights are extracted as well.

The rest of this paper is presented as follows. Section 2 briefly reviews studies regarding production scheduling in HMRSs and RMSs, and emphasizes the novel aspects of this study. In Section 3, the investigated scheduling problem in an HMRS is elaborated. Section 4 provides the MILP model and its equivalent CP model. The WMILP and the LBBD approach are expressed in Section 5. Section 6 focuses on evaluating the proposed solution methods, in addition to performing relevant sensitivity analysis on the impactful parameters of the problem. Finally, Section 7 summarizes the study and proposes some directions for further research on this topic.

## 2. Literature review

In this section, the related scheduling problems in two areas, including HMRSs and RMSs, are briefly reviewed, and at the end, the contributions of this study are accentuated.

### 2.1. Scheduling in hybrid manufacturing-remanufacturing systems

The literature regarding the scheduling in dedicated remanufacturing systems is well-established. Some studies have only focused on the disassembly stage (e.g., Ehm [21]) or just processing of EOL products (e.g., Shi et al. [22]), while others considered scheduling decisions related to disassembly, reprocessing, and assembly in an integrated manner (e.g., Wang et al. [23]; Guo et al. [24]). On the contrary, there is a few research regarding scheduling in hybrid manufacturing-remanufacturing systems.

Regarding a stable demand pattern over the long term, Tang and Teunter [25] studied a multi-product economic lot scheduling problem in which, for each product, there is one lot for manufacturing and one lot for remanufacturing. Considering a common cycle policy, they provided an MILP-based algorithm to determine the optimal lot size and their sequence, minimizing the sum of the inventory holding cost and sequence-independent setup cost. Due to the impracticality of this algorithm in large-sized problems, Teunter, Tang and Kaparis [26] later developed a heuristic method capable of finding near-optimal solutions. Zanoni et al. [27] further extended this problem by relaxing the common cycle restriction and the presence of only one lot for (re)manufacturing, and presented a basic period policy to solve it. Aminipour, Bahroun and Hariga [28] delved into a cyclic scheduling problem with a cyclic pattern in each product demand. The demand can be fulfilled by both manufactured and remanufactured products. The cycle is discretized into multiple periods in which only the manufacturing or remanufacturing of one product can be performed. They proposed an MILP model for the problem to minimize the sum of manufacturing and remanufacturing setup cost, and holding cost for the EOL items, manufactured and remanufactured products.

Concerning a variable demand pattern over short- or mid-term, an integrated lot-sizing and scheduling problem (ILSP) was explored by Giglio et al. [29] in a job shop environment with compressible processing times. They assumed that manufactured and remanufactured products are the same in satisfying the demand. Moreover, the processing routes for manufacturing and remanufacturing products are the same, with different nominal processing times. They proposed a Mixed-Integer programming (MIP) model and a relax-and-fix heuristic for the problem considering the total cost of energy, back-logging, setup, production, and EOL and serviceable products inventory. Later, Roshani et al. [30] considered similar settings for an ILSP problem on a single machine problem besides the incorporation of sequence-dependent setup times and costs for both manufacturing and remanufacturing of products. They presented an MILP model, a relax-and-fix heuristic, and a grey wolf optimization algorithm to tackle the complexity of the problem. Torkaman et al. [31] studied an ILSP in a flow shop environment with sequence-dependent setup time/cost between two different types of products. Moreover, different quality levels for EOL products influence the processing times for remanufacturing. In addition, no setup was considered between the manufacturing and remanufacturing a specific product. They developed an MIP, a rolling horizon heuristic, and a hybrid metaheuristic to solve the problem, optimizing the total cost of inventory (EOL product, work in progress, and final products), setup, manufacturing, and remanufacturing.

### 2.2. Scheduling in reconfigurable manufacturing systems

The Reconfigurable Manufacturing System, as a novel manufacturing paradigm, has demonstrated a significant performance in confronting current manufacturing challenges [32, 33]. It has a customized flexibility that enables it to adjust to production requirements efficiently [18, 19]. So far, different problems arising from the management of RMSs, like system design, process planning, production planning, and production scheduling, have been investigated [19, 33]. Nevertheless, research related to RMS scheduling is still in its early stages.

In RMS scheduling, besides scheduling jobs on machines, decisions regarding the reconfigurations of the system and machines during the planning horizon are necessary. Accordingly, system-level reconfiguration (i.e., adding, removing, or rearranging machines and equipment) has been discovered in cellular manufacturing [34], flow line [35], flexible transfer line [36], and assembly line [37]. In addition, machine-level reconfiguration (i.e., adding, removing, or modifying modules of machines) has been incorporated in the scheduling of single machine [38], parallel machine [39], flow shop [40, 41], job shop [42, 43], flexible job shop [44, 45] environments. Other production decisions have also been integrated into RMS scheduling, including process planning [46, 47], workforce planning [39, 42], lot-sizing [48, 49], outsourcing [40], and machine module allocation [50].

Due to the inclusion of the above-mentioned decisions in RMS scheduling problems, they bear higher complexity than conventional production scheduling problems. As a result, in addition to mathematical models, different solution approaches have been developed to confront these complexities, including CP [42], metaheuristic (e.g., memetic algorithm [38], multi-objective particle swarm optimization algorithm [35], differential evolution algorithm [44], equilibrium optimizer [51], Non-dominated Sorting Genetic Algorithm III [46], genetic algorithm [50], and multi-start evolutionary local search [43]), heuristic [47], machine learning (e.g., deep reinforced learning [52]) and hybrid methods (e.g., a combination of MILP and CP models [39], and a combination of genetic algorithm and MILP model [49]).

### 2.3. Contributions of this study

Although Aljuneidi and Bulgak [16] and Aljuneidi and Bulgak [17] investigated the system-level reconfigurations in HMRSs at tactical production decision-making (i.e., production planning), no research has studied the machine-level reconfigurations in operational decision-making (i.e., production scheduling) of HMRSs. Moreover, despite the benefit of batch processing in many practical situations, like additive manufacturing in reducing unnecessary setups [53], it has been explored in neither HMRS scheduling nor RMS scheduling problems. To fill these gaps, this study provides the following contributions to the literature:

1). Focusing on the scheduling of an HMRS benefiting from RMS concepts to facilitate the inclusion of manufacturing and remanufacturing in a shared production system. In this regard, the machines (stations or cells) can be reconfigured to provide the required production capabilities as required.

2). Considering the possibility of processing manufacturing or remanufacturing orders in separate batches, or combining them in a same batch, for scheduling in a reconfigurable HMRS.

3). Developing novel MILP and CP models to formally formulate the investigated problem.

4). Devising an efficient LBBD to confront the complexity of the problem. Accordingly, the master problem deals with assigning orders to machine-configuration combinations, and sub-problems involve batching and scheduling orders for each machine.

5). Employing the warm start technique to boost the performance of the developed MILP model (namely WMILP) using a two-step procedure consisting of mathematical models.

6). Performing comprehensive computational experiments to assess the performance of the proposed MILP model, CP model, LBBD method, and WMILP, as well as providing relevant managerial insights.

Contributions 1 and 2 refer to the knowledge this paper generates on the scheduling of HMRSs. We explicate how the results advance the relevant literature and extend existing knowledge of HMRS. Contributions 3-6 refer to innovative methodologies, including the MILP model, CP model, LBBD method, and WMILP. For example, the underlying challenge of the LBBD is that the execution of the method might be halted due to time and memory limitations and the presence of many inactive cuts, which result in a large MP with a high solution time. As investigated by Rahmaniani et al. [54], nearly 90% of the overall running time in these types of decomposition methods is used for solving the MP. In the present research, our decomposition strategy specifically focuses on finding a way to partition the problem effectively into MP and SPs. This involves developing new properties that account for problem-specific characteristics and exploring alternative ways to divide the LBBD into MP and SPs. Another significant contribution of the LBBD to overcome such a challenge is to find a tight lower bound (LB) that is a relaxation of the SPs.

We focus on a situation where the aforementioned exact methods can be more suitable than metaheuristics for scheduling HMRSs. A reason behind this intuition is that metaheuristics, while capable of exploring a larger solution space, often do so at the cost of greater variability (e.g., see [43] for the case of RMSs). Accordingly, the study is limited to computational experiments involving the MILP model, CP model, LBBD method, and WMILP.

## 3. Problem description

We first present the notation for the scheduling problem, followed by an explanation of the components of a hybrid manufacturing-remanufacturing system. Table 1 summarizes the key notation and symbols, with detailed definitions provided throughout the section. The section elaborates on the scheduling problem regarding a hybrid manufacturing-remanufacturing system comprising a set of parallel non-identical reconfigurable machines (stations or production cells) denoted by $\mathcal{M}$. The set of orders is represented by $\mathcal{O}$. More precisely, two types of orders are received in the system. Manufacturing orders that are produced by raw materials, and remanufacturing orders that require the processing of EOL products. We denote as disjoint sets $M\mathcal{O}$ and $R\mathcal{O}$ the sets of manufacturing and remanufacturing orders, respectively. As a result, different production capabilities are necessary to meet the demands of this shared system provided by the reconfigurable machines. Assume machine $m$ with configuration $c$ shows a machine-configuration combination. Then, each machine configuration may satisfy the requirements of a predefined number of manufacturing or remanufacturing orders ($o \epsilon \mathcal{O}$), which is called the eligible configuration ($\mathcal{E}_{omc} = 1$). On the other hand, at least one eligible machine-configuration combination is capable of processing an order. In order to save setup times, it is possible to group parts into batches under specific machine configurations. In this case, the setup time ($BST_{mc}$) is shared by the orders of a particular batch $b$. However, each machine $m$ has a limited processing area capacity of $MPA_m$ and height of $MPH_m$ that constrain the inclusion of orders in a batch. The other assumptions of this problem are listed as follows.

- The set of orders $\mathcal{O}$ and their requirements are known and available at the beginning of the planning horizon.

- The processing time ($OPT_{omc}$) of an order is deterministic and dependent on the assigned machine-configuration combination.
- Based on the requirements of orders, a batch on a specific configuration of a machine can include only manufacturing orders from set $MO$, only remanufacturing orders from set $RO$, or both manufacturing and remanufacturing orders.
- Each order height and the total area occupied by the orders of a batch are denoted by $OH_o$ and $OA_o$, respectively, and cannot surpass the processing height $MPH_m$ and area limits $MPA_m$ of its related machine $m$.
- The processing time of a batch is the total processing times of its assigned orders following the definition of serial-batch processing.
- Each batch is performed by a configuration of its corresponding machine, and its setup time depends on the assigned configuration.
- At any moment, at most, one batch can be processed on a machine.
- During the processing of a batch, it cannot be interrupted, and any order cannot be added or removed from it.
- If the selected configurations of two adjacent batches on a machine $m$ are different, a reconfiguration time of $\mathcal{T}_{mcc'}$ is required depending on the former and latter configurations of $c$ and $c'$. Moreover, the reconfiguration times follow the triangle inequality, i.e, $\mathcal{T}_{mcc'} + \mathcal{T}_{mc'c''} \geq \mathcal{T}_{mcc''} \quad \forall m \in \mathcal{M}, \forall c, c', c'' \in \mathcal{C}_m$.
- During a reconfiguration, no batch can be processed on the corresponding machine.

**Table 1.** Important notation and symbols

| | |
|---|---|
| **Sets and indices** | |
| $\mathcal{MO}$ | Set of manufacturing orders |
| $\mathcal{RO}$ | Set of remanufacturing orders |
| $\mathcal{O}$ | Set of orders, $\mathcal{O} = \mathcal{MO} \cup \mathcal{RO}$ |
| $\mathcal{B}$ | Set of batches |
| $\mathcal{M}$ | Set of machines |
| $\mathcal{C}_m$ | Set of configurations belongs to machine $m$ |
| $o, o'$ | Index of order |
| $b$ | Index of batch |
| $m$ | Index of machine |
| $c$ | Index of configuration; $c = 0$ represents the initial configuration |
| **Parameters** | |
| $OA_o$ | Order area: the area occupied by order $o$ in a batch |
| $OH_o$ | Order height: the height of order $o$ |
| $MPA_m$ | Machine processing area: the maximum area provided by machine $m$ for batch processing |
| $MPH_m$ | Machine processing height: the maximum acceptable height of an order to be processed by machine $m$ |
| $OPT_{omc}$ | Order processing time: the time required to process order $o$ by configuration $c$ of machine $m$ |
| $BST_{mc}$ | Batch setup time on machine $m$ with configuration $c$ |
| $BL_{mc}$ | Batching limit: 1 if at most one order can be placed within a batch on machine $m$ with configuration $c$; 0, otherwise |
| $\mathcal{E}_{omc}$ | 1 if configuration $c$ of machine $m$ is eligible to process order $o$ |
| $\mathcal{T}_{mcc'}$ | The time needed to reconfigure machine $m$ from configuration $c$ to $c'$ |
| $G$ | A large positive number |
| **Variables** | |
| $x_{obmc}$ | 1 if order $o$ is assigned to batch $b$ of machine $m$ being processed with configuration $c$; 0, otherwise |
| $y_{bmc}$ | 1 if batch $b$ of machine $m$ is formed and being processed with configuration $c$; 0, otherwise |
| $z_{mcc'}$ | 1 if batches with configuration $c'$ are processed after those with configuration $c$; 0, otherwise |
| $w_{mc}$ | 1 if configuration $c$ of machine $m$ is utilized; 0, otherwise |
| $cd_{mc}$ | Configuration duration: the time required to process batches with configuration $c$ on machine $m$ including their setup times |
| $cct_{mc}$ | Configuration completion time: the completion time of batches with configuration $c$ on machine $m$ |
| $C_{max}$ | The maximum completion time of all configurations or batches (i.e., makespan) |

The problem seeks simultaneous decision-making concerning (i) the assignment of each order to an eligible machine-configuration respecting the height and area limits, (ii) batching the orders assigned to each machine-configuration respecting the area limit, and (iii) the scheduling of the batches on the machines to minimize the makespan. Fig. 2 illustrates the problem by a simple example. As shown in the figure, there are 2 reconfigurable machines in the system, each with 3 configurations. The assignment of orders to machine-configurations, batching of orders on machines, and their scheduling are also presented. For instance, remanufacturing order $O_1$ and manufacturing order $O_2$ are assigned to $Configuration_1$ and $Configuration_2$ of $Machine_1$, respectively. Manufacturing orders $O_2$, $O_3$, and $O_4$ are included in $Batch_1$ of $Machine_1$, and remanufacturing orders $O_1$ and $O_5$ on $Batch_2$ $Machine_1$. A reconfiguration is required between the processing of $Batch_1$ and $Batch_2$ of $Machine_1$ from $Configuration_2$ to $Configuration_1$ ($C_2$-$C_1$). Moreover, due to the similar processing requirements of manufacturing order $O_7$ and remanufacturing order $O_8$, it is possible to process them within $Batch_1$ of $Machine_2$.

## 4. Problem formulation

In this section, first, two properties of an optimal solution to the reconfigurable HMRS scheduling problem are introduced. Afterward, the notation of the reconfigurable HMRS scheduling problem is presented. Subsequently, the MILP and CP models are extended.

### 4.1. Properties of an optimal solution to the problem

Prior to presenting the MILP model, we first introduce two properties of an optimal solution to the reconfigurable HMRS scheduling problem. Employing them in the modeling leads to a reduction in the solution space and enhances its performance accordingly. These properties are a modification of the ones provided by Vahedi-Nouri et al. [39] for workforce planning and scheduling in a parallel machine environment consisting of reconfigurable machines.

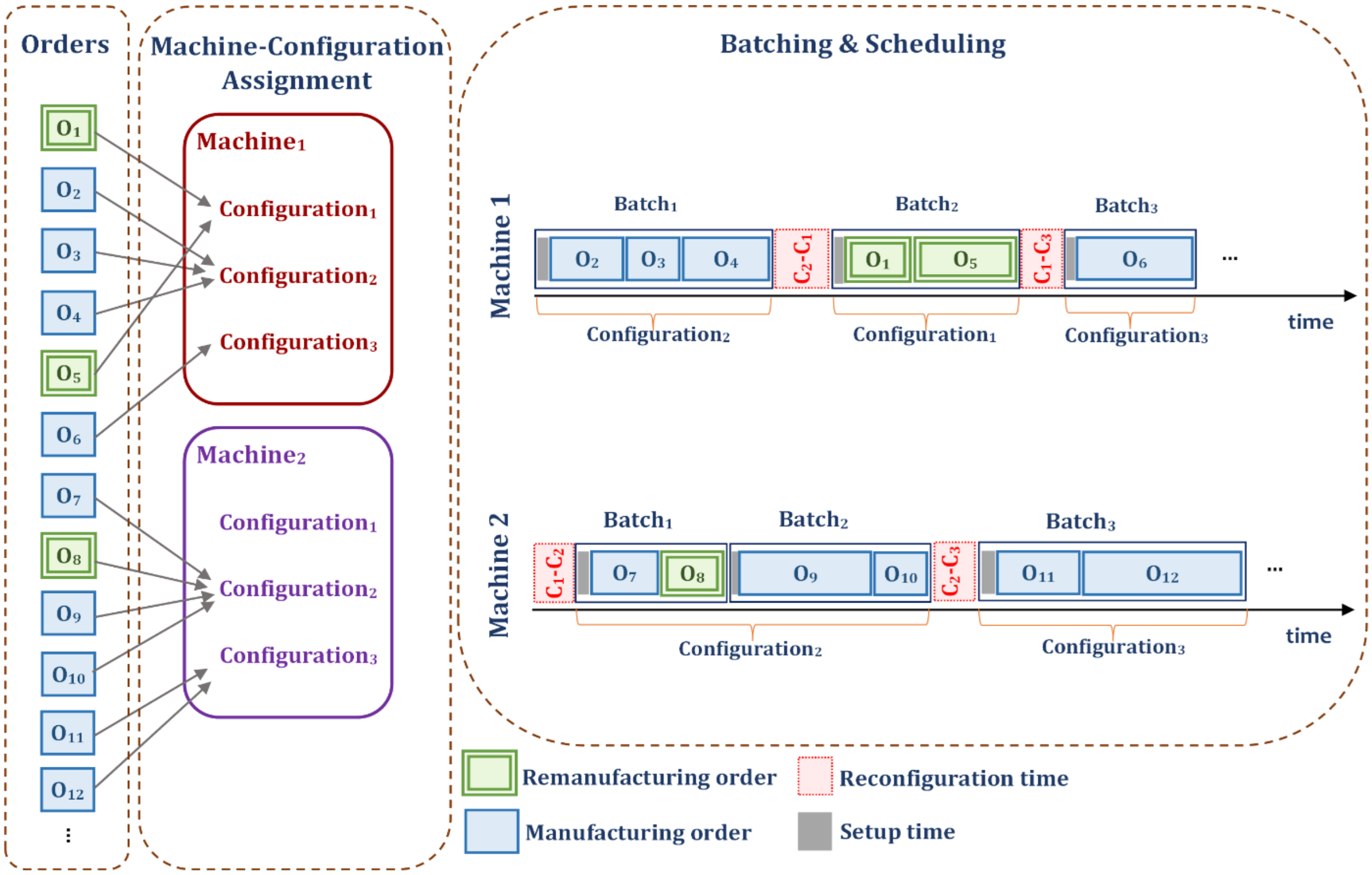

**Fig. 2.** A simple illustrative example of the reconfigurable HMRS scheduling problem.

**Property I.** *In the case of the reconfigurable HMRS scheduling problem, there is always an optimal solution where, on every machine, all the batches with an identical configuration are processed successively.*

**Proof.** Two following cases are needed to be explored in this regard:

**1)** *For machine m in an optimal solution in which the completion time of its last batch is equal to $C_{max}$*: following the proof by contraction method, assume $S_m^1$ represents the optimal schedule on machine $m$ in which Property I is not held. Thus, at least one batch can be found processed apart from other batches with the same configuration on machine $m$. Fig. 3(a) demonstrates $S_m^1$, in which $RT_{[k]}$ denotes the reconfiguration time, $B_{[r]}$ the batch on the $r^{th}$position of $\mathcal{S}_m^1$, and each color signifies a configuration of machine $m$. The makespan of $\mathcal{S}_m^1$ ($C_{max}^{\mathcal{S}_m^1}$) is determined by Equations (1) and (2), where $c_{[1]}$ is the first employed configuration in $\mathcal{S}_m^1$, and $bpt_{bm}$ is the processing time of a batch including its setup time.

$$C_{max}^{\mathcal{S}_m^1} = RT_{[1]} + \sum_{b \in \{B_{[1]},\dots,B_{[r]}\}} bpt_{bm} + RT_{[2]} + \ell + RT_{[k]} + bpt_{B_{[s]},m} \tag{1}$$

$$bpt_{bm} = BST_{m,c_{[1]}} + \sum_{o \in O} OPT_{omc_{[1]}} \cdot x_{obmc_{[1]}}, \qquad \forall b \in \{B_{[1]}, \dots, B_{[r]}\} \tag{2}$$

Considering the fact that no precedence relationship exists between orders, a feasible solution can be reached by inserting $B_{[s]}$ just after $B_{[r]}$ without any needs of reconfiguration. This solution, referred to as $\mathcal{S}_m^2$, is shown in Fig. 3(b). Hence, $C_{max}^{\mathcal{S}_m^2}$ can be calculated by Equation (3):

$$C_{max}^{\mathcal{S}_m^2} = RT_{[1]} + \sum_{b \in \{B_{[1]},\dots,B_{[r]}\}} bpt_{bm} + bpt_{B_{[s]},m} + RT_{[2]} + \ell \tag{3}$$

Comparing the makespan of the two schedules, it is evident that $C_{max}^{\mathcal{S}_m^1} - C_{max}^{\mathcal{S}_m^2} = RT_{[k]} > 0$. Hence, the makespan of $\mathcal{S}_m^2$ is lower than of $\mathcal{S}_m^1$, which is in contradiction to the assumption. In the same way, if more than one batch does not observe Property I, this difference can be even enlarged.

**2)** *For machine m in an optimal solution in which the completion time of its last batch is less than* $C_{max}$: assume that Fig. 3(a) indicates this case. Thus, by processing $B_{[s]}$ just after $B_{[r]}$ (as shown in Fig. 3(b)), $C_{max}^{\mathcal{S}_m^1}$ is decreased. Hence, it does not deteriorate $C_{max}$. As a result, Property I is held, and the proof is completed. □

**Property II.** *Considering an optimal solution to the reconfigurable HMRS scheduling problem complied with Property I, various sequences of batches processed with an identical configuration result in an equal* $C_{max}$*.*

**Proof.** Since no reconfiguration is required between processing successive batches of a machine with an identical configuration, altering their positions in the sequence does not impact $C_{max}$. For instance, assume Fig. 3(b) expresses an optimal schedule $\mathcal{S}_m^2$. Swapping $B_{[r]}$ with $B_{[s]}$, leads to schedule $\mathcal{S}_m^2$ (Fig. 3(c)) with the same makespan. □

By employing these properties, instead of scheduling batches individually, we just need to schedule utilized configurations on each machine. For instance, if 10 batches are assigned to be processed by 3 configurations of a machine, instead of having 10! possibilities for the sequencing, the model just explores 3! possibilities. As a result, the performance of the models is significantly improved.

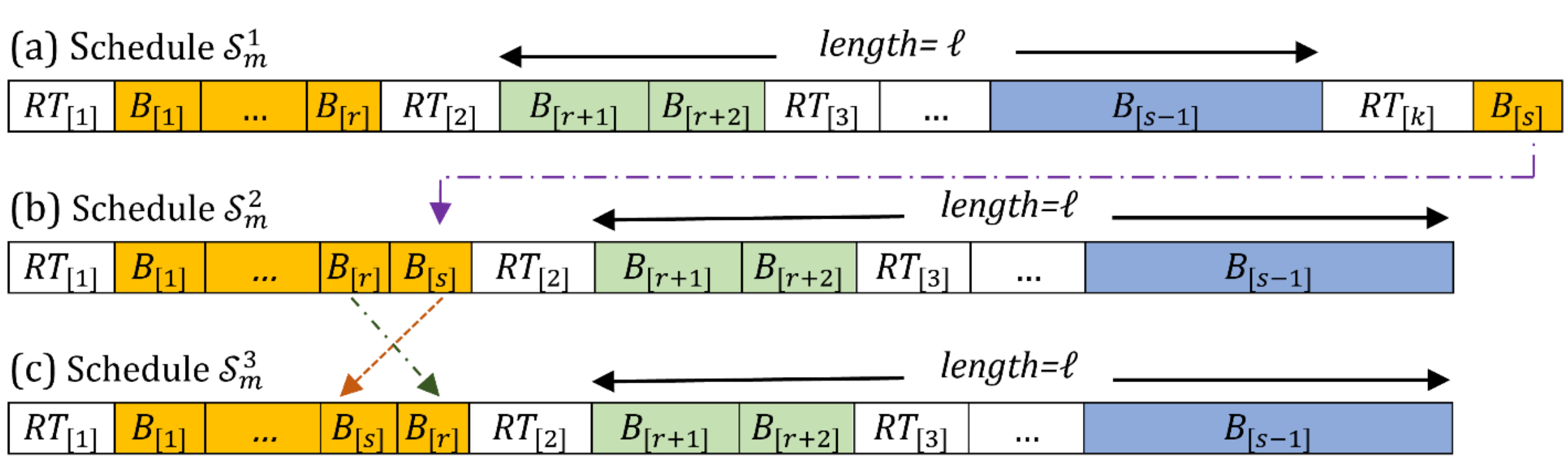

**Fig. 3.** Three schedules of a machine demonstrating properties of an optimal solution

### 4.2. MILP model

$$Minimize\ C_{max} \tag{4}$$

Subject to:

$$\sum_{b\in\mathcal{B}}\sum_{m\in\mathcal{M}}\sum_{c\in\mathcal{C}_m|\mathcal{E}_{omc}=1} x_{obmc} = 1, \quad \forall o \in \mathcal{O} \tag{5}$$

$$\sum_{o\in\mathcal{O}|\mathcal{E}_{omc}=1} x_{obmc} \leq 1, \quad \forall b \in \mathcal{B}, m \in \mathcal{M}, \forall c \in \mathcal{C}_m | BL_{mc} = 1 \tag{6}$$

$$\sum_{c\in\mathcal{C}_m} y_{bmc} \leq 1, \quad \forall b \in \mathcal{B}, m \in \mathcal{M} \tag{7}$$

$$\sum_{o\in\mathcal{O}|\mathcal{E}_{omc}=1} x_{obmc} \leq |\mathcal{O}| \cdot y_{bmc}, \quad \forall b \in \mathcal{B}, m \in \mathcal{M}, \forall c \in \mathcal{C}_m \tag{8}$$

$$\sum_{b\in\mathcal{B}}\sum_{c\in\mathcal{C}_m} x_{obmc} \cdot OH_o \leq MPH_m, \quad \forall o \in \mathcal{O}, m \in \mathcal{M} \tag{9}$$

$$\sum_{o\in\mathcal{O}}\sum_{c\in\mathcal{C}_m|\mathcal{E}_{omc}=1} x_{obmc} \cdot OA_o \leq MPA_m, \quad \forall b \in \mathcal{B}, m \in \mathcal{M} \tag{10}$$

$$cd_{mc} = \sum_{o\in\mathcal{O}|\mathcal{E}_{omc}=1}\sum_{b\in\mathcal{B}} x_{obmc} \cdot OPT_{omc} + BST_{mc} \cdot \sum_{b\in\mathcal{B}} y_{bmc}, \qquad \forall m \in \mathcal{M}, \forall c \in \mathcal{C}_m \tag{11}$$

$$\sum_{b\in\mathcal{B}} y_{bmc} \leq |\mathcal{B}| \cdot w_{mc}, \qquad \forall m \in \mathcal{M}, \forall c \in \mathcal{C}_m \tag{12}$$

$$cct_{mc} \geq cd_{mc} + \mathcal{T}_{m0c} - G \cdot (1 - w_{mc}), \qquad \forall m \in \mathcal{M}, \forall c \in \mathcal{C}_m \tag{13}$$

$$cct_{mc'} \geq cct_{mc} + cd_{mc'} + \mathcal{T}_{mcc'} - G \cdot (3 - z_{mcc'} - w_{mc} - w_{mc'}), \tag{14}$$
$$\forall m \in \mathcal{M}, \forall c, c' \in \mathcal{C}_m | c \neq c'$$

$$z_{mcc'} + z_{mc'c} \geq w_{mc} + w_{mc'} - 1, \qquad \forall m \in \mathcal{M}, \forall c, c' \in \mathcal{C}_m | c \neq c' \tag{15}$$

$$z_{mcc'} + z_{mc'c} \leq 3 - w_{mc} - w_{mc'}, \qquad \forall m \in \mathcal{M}, \forall c, c' \in \mathcal{C}_m | c \neq c' \tag{16}$$

$$C_{max} \geq cct_{mc}, \qquad \forall m \in \mathcal{M}, \forall c \in \mathcal{C}_m \tag{17}$$

$$x_{obmc}, y_{bmc}, z_{mc'c}, w_{mc} \in \{0,1\};\ C_{max}, cct_{mc}, cd_{mc} \in \mathbb{R}^+, \tag{18}$$

$$\forall o \in \mathcal{O}, \forall b \in \mathcal{B}, m \in \mathcal{M}, \forall c, c' \in \mathcal{C}_m$$

Expression (4) indicates the objective function, i.e., makespan. Constraint (5) is related to assigning each order to a batch-machine-configuration combination respecting the configuration eligibility. Constraint (6) restricts the number of orders in a batch of a machine with a specific configuration to at most one item if its related batch processing limit equals 1. Constraint (7) concerns the formation of a batch on a machine and determines its configuration. Constraint (8) explains that an order can be assigned to a batch-machine-configuration combination if the related batch is formed. Machine processing height and area limits are respected in Constraints (9) and (10), respectively. Constraint (11) calculates the duration of each configuration on a machine based on the orders assigned to the relevant batches and required batch setup times. Constraint (12) relates that a batch on a machine with a specific configuration is formed if the corresponding configuration of the machine is utilized. Constraints (13) and (14) impose the required reconfiguration times from the initial configuration, and between utilized configurations on each machine, respectively. Moreover, they determine the completion times of configurations. The sequence of utilized configurations on each machine is specified in Constraints (15) and (16). Constraint (17) concerns the makespan. Constraint (18) defines the domains of the introduced variables.

**4.3. CP model**

In recent years, Constraint Programming has demonstrated itself as a competitive solution approach to tackle combinatorial optimization problems, specifically scheduling problems (e.g., [55-57]). It enjoys specific variables and functions with which a vast scope of problems can be effectively formulated.

In this paper, we adopt IBM ILOG CP Optimizer (CPO) to formulate the CP model. One of the main features of CP is the *interval* variable representing an activity. In CPO, the start, end, and duration of an interval can be recalled by $startOf$(), $endOf$(), and $sizeOf$() functions. Moreover, if an interval is defined as optional, it can be present or absent in the schedule, which can be recalled by $presenceOf$ function. In order to be familiarized with the basic information regarding the variables and functions in the CPO, you can refer to Laborie, Rogerie [58]. In the following, first, the relevant CP variables of the problem are defined, then, the CP model is formulated.

*Variables*

| | |
|---|---|
| $OAsn_{obmc}$ | An optional interval variable signifying the assignment of order $o$ to batch $b$ of machine $m$ being processed with configuration $c$ |
| $Bch_{bmc}$ | An optional interval variable signifying batch $b$ of machine $m$ being processed by configuration $c$ |
| $Cnf_{mc}$ | An optional interval variable signifying configuration $c$ of machine $m$ |
| $Sq_m$ | A sequence variable representing the sequence of configurations on machine $m$ in which index $c$ is tracked by *types option* for imposing reconfiguration times (i.e., $Sq_m\ (Cnf_{mc}, types$ c, $\forall c \in \mathcal{C}_m)$) |

$$Minimize\ C_{max} = \max_{m\in\mathcal{M},c\in\mathcal{C}_m} endOf(Cnf_{mc}) \tag{19}$$

Subject to:

$$\sum_{b\in\mathcal{B}}\sum_{m\in\mathcal{M}}\sum_{c\in\mathcal{C}_m|\mathcal{E}_{omc}=1} presenceOf(OAsn_{obmc}) = 1, \quad \forall o \in \mathcal{O} \tag{20}$$

$$\sum_{o\in\mathcal{O}|\mathcal{E}_{omc}=1} presenceOf(OAsn_{obmc}) \leq 1, \quad \forall b \in \mathcal{B}, m \in \mathcal{M}, \forall c \in \mathcal{C}_m|BL_{mc} = 1 \tag{21}$$

$$\sum_{c\in\mathcal{C}_m} presenceOf(Bch_{bmc}) \leq 1, \quad \forall b \in \mathcal{B}, m \in \mathcal{M} \tag{22}$$

$$\sum_{o\in\mathcal{O}|\mathcal{E}_{omc}=1} presenceOf(OAsn_{obmc}) \leq |\mathcal{O}| \cdot presenceOf(Bch_{bmc}), \tag{23}$$

$$\forall b \in \mathcal{B}, \forall m \in \mathcal{M}, \forall c \in \mathcal{C}_m$$

$$\sum_{b\in\mathcal{B}}\sum_{c\in\mathcal{C}_m} presenceOf(OAsn_{obmc}) \cdot OH_o \leq MH_m, \quad \forall o \in \mathcal{O}, m \in \mathcal{M} \tag{24}$$

$$\sum_{o\in\mathcal{O}}\sum_{c\in\mathcal{C}_m|\mathcal{E}_{omc}=1} presenceOf(OAsn_{obmc}) \cdot OA_o \leq MA_m, \quad \forall b \in \mathcal{B}, m \in \mathcal{M} \tag{25}$$

$$sizeOf(Bch_{bmc}) = \sum_{o\in\mathcal{O}|\mathcal{E}_{omc}=1} presenceOf(OAsn_{obmc}) \cdot OPT_{omc} + \tag{26}$$

$$BST_{mc} \cdot presenceOf(Bch_{bmc}), \qquad \forall b \in \mathcal{B}, \forall m \in \mathcal{M}, \forall c \in \mathcal{C}_m$$

$$noOverlap([Bch_{bmc}]_{\forall b\in\mathcal{B}}), \qquad \forall m \in \mathcal{M}, \forall c \in \mathcal{C}_m \tag{27}$$

$$span(Cnf_{mc}, [Bch_{bmc}]_{\forall b\in\mathcal{B}}), \qquad \forall m \in \mathcal{M}, \forall c \in \mathcal{C}_m \tag{28}$$

$$startOf(Cnf_{mc}) \geq \mathcal{T}_{m0c} \cdot presenceOf(Cnf_{mc}), \qquad \forall m \in \mathcal{M}, \forall c \in \mathcal{C}_m \tag{29}$$

$$noOverlap(Sq_m, R_m), \qquad \forall m \in \mathcal{M} \tag{30}$$

Equation (19) is the objective function of the CP model minimizing the makespan. Order assignment is observed by Constraint (20) considering the machine-configuration eligibility. Constraint (21) respects the batching limit restriction. Configuration of a batch of a machine is specified by Constraint (22). Constraint (23) indicates that the assignment of an order to a batch-machine combination depends on the presence of the relevant batch interval. Constraints (24) and (25) observe the machine processing height and area limits, respectively. Constraint (26) determines the size of each batch interval ($Bch_{bmc}$) based on its assigned orders and the required setup time. Applying $noOverlap()$ function, Constraint (27) ensures that there is no overlap among batch intervals with a same configuration. Constraint (28) expresses that a present configuration of a machine starts and ends with the first and last batches of the machine with this configuration, respectively. Constraint (29) imposes the reconfiguration time from the initial configuration of a machine if it is required. Constraint (30) applies $noOverlap()$ function on sequence $Sq_m$ that guarantees that there is no overlap among its present intervals ($Cnf_{mc}$) on the machine. In addition, it respects reconfiguration times between the consecutive configurations of each machine specified in transition matrix $R_m$.

## 5. Solution Methods

In the following subsections, the developed LBBD method and WMILP model are illustrated, respectively.

### 5.1. Logic-based Benders decomposition

LBBD is a known exact solution method for tackling complex combinatorial optimization problems [59]. Similar to Benders decomposition (BD), LBBD decomposes the original model into one master problem (MP) and one or multiple subproblems (SPs). However, LBBD rectifies BD's shortcomings in handling non-linear, integer, and mixed-integer SPs [60]. LBBD iterates between MP and SP(s). In minimization problems, at each iteration, the MP solution represents a lower bound (LB) for the original model, and it is used to generate SP(s). Besides, SP(s) contribute to the best upper bound (UB) and feasibility and optimality cuts that are added to the MP to remove previously reached solutions. LBBD terminates when MP and SP(s) reach a same objective, or a predefined termination condition is achieved (e.g., reaching a specific gap, or computational time or iteration limit) [60, 61]. Unlike BD, LBBD does not employ a conventional procedure to generate cuts. Instead, the cuts must be developed based on the structure of an investigating problem[62].

LBBD has been successfully applied in a variety of problems, including maximal covering location problem [63], railway scheduling [64], electric vehicle routing problem [65], integrated process configuration and production planning [66], shelf space allocation [67], integrated parallel machine scheduling and location problem [68], integrated lot-sizing and scheduling problem [57, 69], operating room scheduling [70], to name a few. In the following, the decomposition scheme of the reconfigurable HMRS scheduling problem, the relevant MP, SPs, optimality cut, and LBBD improvement are elaborated.

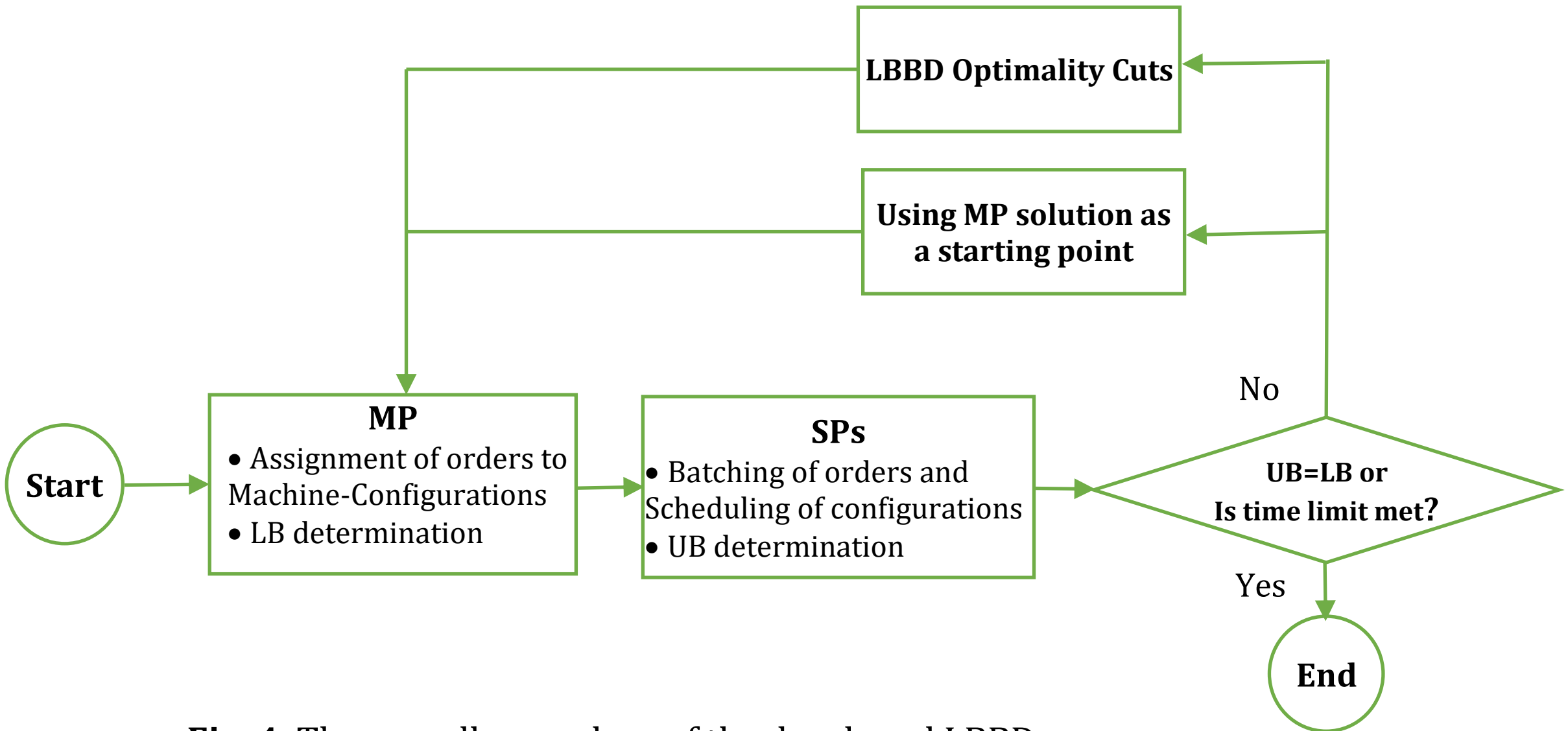

**Fig. 4.** The overall procedure of the developed LBBD.

### 5.1.1. Problem decomposition

The overall procedure of the developed LBBD is presented in Fig. 4. As can be seen, the MP is responsible for determining the assignment of each order to an eligible machine-configuration combination. On the other hand, each SPs is involved in batching the assigned orders and scheduling the batches on their relevant machines. As a result, there are one MP and $|\mathcal{M}|$ independent SPs. It should be noted that the SPs are always feasible since there is no limit on the number of assigned orders to a machine or the completion time of a machine. If the maximum makespan of all SPs is equal to the MP's objective, then the solution is optimal, and the algorithm is terminated. Otherwise, it is sub-optimal. In this case, if the termination condition (e.g., computational time limit) is not met, for each SP with the makespan equal to the maximum makespan, an optimality cut is extended and added to the MP, and the algorithm proceeds. Each part of the LBBD procedure is detailed in the next sub-sections.

#### 5.1.2. Master problem

The MP assigns orders to machine-configuration combinations, leaving orders batching and configurations scheduling on each machine to the SPs. Accordingly, the MP is a relaxation of the original model in which $x_{obmc}$ is transformed to $x_{omc}$. Moreover, $z_{mcc'}$ (related to the sequencing of configurations), $y_{bmc}$, $cd_{mc}$, and $cct_{mc}$ are removed. The additional parameter and variables of the MP are as follows.

$MRT_{mc}$ The minimum reconfiguration time to configuration $c$ of machine $m$ from its other configurations, including initial configuration.

$x_{omc}$ 1 if order $o$ is assigned to machine $m$ being processed with configuration $c$; 0, otherwise.

$ct_m$ Completion time of machine $m$.

$$Minimize\ Obj \tag{31}$$

Subject to:

$$\sum_{m\in\mathcal{M}} \sum_{c\in\mathcal{C}_m|\mathcal{E}_{omc}=1} x_{omc} = 1, \quad \forall o \in \mathcal{O} \tag{32}$$

$$\sum_{o\in\mathcal{O}|\mathcal{E}_{omc}=1} x_{omc} \leq |\mathcal{O}| \cdot w_{mc}, \quad m \in \mathcal{M}, \forall c \in \mathcal{C}_m \tag{33}$$

$$\sum_{c\in\mathcal{C}_m} x_{omc} \cdot OH_o \leq MPH_m, \quad \forall o \in \mathcal{O}, m \in \mathcal{M} \tag{34}$$

$$\sum_{c\in\mathcal{C}_m} x_{omc} \cdot OA_o \leq MPA_m, \quad \forall o \in \mathcal{O}, m \in \mathcal{M} \tag{35}$$

$$ct_m = \sum_{o \in \mathcal{O}} \sum_{c \in \mathcal{C}_m | \mathcal{E}_{omc}=1} x_{omc} \cdot OPT_{omc} + \sum_{c \in \mathcal{C}_m} MRT_{mc} \cdot w_{mc} + \sum_{c \in \mathcal{C}_m | BL_{mc}=0} BST_{mc} \cdot \left( \left(\sum_{o \in \mathcal{O} | \mathcal{E}_{omc}=1} x_{omc} \cdot OA_o\right) \Big/ MPA_m \right) + \sum_{c \in \mathcal{C}_m | BL_{mc}=1} BST_{mc} \cdot \sum_{o \in \mathcal{O} | \mathcal{E}_{omc}=1} x_{omc}, \quad \forall m \in \mathcal{M} \tag{36}$$

$$Obj \geq ct_m, \quad \forall m \in \mathcal{M} \tag{37}$$

$$\text{Benders optimality cuts} \tag{38}$$

$$x_{omc}, u_{mcc'}, w_{mc} \in \{0,1\};\ Obj, ct_m \in \mathbb{R}^+, \quad \forall o \in \mathcal{O}, \forall m \in \mathcal{M}, \forall c, c' \in \mathcal{C}_m \tag{39}$$

Expression (31) presents the objective function, which is an approximation of the makespan. Constraint (32) limits the assignment of each order to an eligible machine-configuration. Constraint (33) indicates that an order can be assigned to a machine-configuration combination if the related configuration is utilized. Machine processing height and are limits are respected in Constraints (34) and (35), respectively. Since the batching and scheduling are not considered in the MP, the true value of the makespan cannot be determined. Incorporating a relaxation of SPs in the MP through a lower bound of SPs' makespan can expedite the LBBD convergence [71]. Accordingly, the completion time of each machine is estimated in Constraint (36), including the processing times of the assigned orders, minimum possible reconfiguration times, and an approximation of batch setup times. Consequently, Constraint (37) concerns the lower bound of the makespan. Constraint (38) indicates the Benders optimality cuts added to the MP during the LBBD iterations. Constraint (39) defines the domains of the variables.

### 5.1.3. Subproblems

Following solving the MP in iteration $h$, an optimal solution with $\{x_{omc}^{h*}, w_{mc}^{h*}\}$ is achieved, determining the assignment of each order to a machine-configuration combination. In this regard, there are $|\mathcal{M}|$ independent subproblems each representing a single machine scheduling problem determining the optimal batching of assigned orders (i.e., for machine $m$, only orders where $x_{omc}^{h*} = 1$) and scheduling of utilized configuration (i.e., $w_{mc}^{h*} = 1$) to minimize the makespan. The SP model for machine $m$ and the relevant variables are as follows.

$x_{obm}^{h}$ 1 if order $o$ is assigned to batch $b$ of machine $m$ in iteration $h$; 0, otherwise

$C_{max}^{hm}$ The makespan of machine $m$ in iteration $h$

$$Minimize\ C_{max}^{hm} \tag{40}$$

Subject to:

$$\sum_{b\in\mathcal{B}} x_{obm}^{h} = 1, \quad \forall o \in \mathcal{O} \middle| \sum_{c\in\mathcal{C}_m} x_{omc}^{h*} = 1 \tag{41}$$

$$\sum_{c\in\mathcal{C}_m | w_{mc}^{h*}=1} y_{bmc}^{h} \leq 1, \quad \forall b \in \mathcal{B} \tag{42}$$

$$\sum_{o\in\mathcal{O}|x_{omc}^{h*}=1} x_{obm}^{h} \leq |\mathcal{O}| \cdot y_{bmc}^{h}, \quad \forall b \in \mathcal{B}, \forall c \in \mathcal{C}_m | w_{mc}^{h*} = 1, BL_{mc} = 0 \tag{43}$$

$$\sum_{o\in\mathcal{O}|x_{omc}^{h*}=1} x_{obm}^{h} \leq y_{bmc}^{h}, \quad \forall b \in \mathcal{B}, \forall c \in \mathcal{C}_m | w_{mc}^{h*} = 1, BL_{mc} = 1 \tag{44}$$

$$\sum_{o\in\mathcal{O}|\sum_{c\in\mathcal{C}_m} x_{omc}^{h*}=1} x_{obm}^{h} \cdot OA_o \leq MPA_m, \quad \forall b \in \mathcal{B} \tag{45}$$

$$cd_{mc}^{h} = \sum_{o\in\mathcal{O}|x_{omc}^{h*}=1} \sum_{b\in\mathcal{B}} x_{obm}^{h} \cdot OPT_{omc} + BST_{mc} \cdot \sum_{b\in\mathcal{B}} y_{bmc}^{h}, \qquad \forall c \in \mathcal{C}_m | w_{mc}^{h*} = 1 \tag{46}$$

$$cct_{mc}^{h} \geq cd_{mc}^{h} + \mathcal{T}_{m0c}, \qquad \forall c \in \mathcal{C}_m | w_{mc}^{h*} = 1 \tag{47}$$

$$cct_{mc'}^{h} \geq cct_{mc}^{h} + cd_{mc'}^{h} + \mathcal{T}_{mcc'} - G \cdot \left(1 - z_{mcc'}^{h}\right), \tag{48}$$

$$\forall m \in \mathcal{M}, \forall c, c' \in \mathcal{C}_m | w_{mc}^{h*} = w_{mc'}^{h*} = 1, c \neq c'$$

$$z_{mcc'}^{h} + z_{mc'c}^{h} = 1, \qquad \forall c, c' \in \mathcal{C}_m | w_{mc}^{h*} = w_{mc'}^{h*} = 1, c \neq c' \tag{49}$$

$$C_{max}^{hm} \geq cct_{mc}^{h}, \qquad \forall c \in \mathcal{C}_m | w_{mc}^{h*} = 1 \tag{50}$$

$$x_{obm}^{h}, y_{bmc}^{h}, z_{mcc'}^{h} \in \{0,1\};\ C_{max}^{hm}, cct_{mc}^{h}, cd_{mc}^{h} \in \mathbb{R}^{+}, \tag{51}$$

$$\forall o \in \mathcal{O} | \sum_{c\in\mathcal{C}_m} x_{omc}^{h*} = 1, \forall b \in \mathcal{B}, \forall c, c' \in \mathcal{C}_m | w_{mc}^{h*} = w_{mc'}^{h*} = 1, c \neq c'$$

Expression (40) represents the makespan of $m$th subproblem (i.e., related to machine $m$) in iteration $h$. Constraints (41)-(44) seek the batching of the assigned orders considering batch processing limit and accordingly determine the configuration of each formed batch. The machine processing area is respected in Constraints (45). Constraint (46) calculates the duration of each utilized configuration on the machine based on the assigned orders to the relevant batches and required batch setup times. Constraints (47)-(49) are related to the scheduling of the configurations incorporating reconfiguration times. Constraint (50) concerns the makespan. Constraint (51) defines the domains of the variables.

#### 5.1.4. LBBD cut

In each iteration of the LBBD algorithm, if the maximum makespan obtained by solving each SP equals the objective of the MP, the optimal solution to the original model is found, and the algorithm terminates. Otherwise, if the termination condition is not met, an optimality cut is added to the MP based on the SPs with the makespan equal to the maximum makespan of all SPs. It should be noted that no feasibility cut is required since any assignment of orders to machine-configuration in the MP leads to feasible solutions to the SPs.

Accordingly, the cut is defined as Constraint (52). Where $C_{max}^{hm*}$ is the makespan found by solving the SP related to machine $m$ in iteration $h$.

$$Obj \geq C_{max}^{hm*} \cdot \left(1 - \sum_{c \in \mathcal{C}_m | w_{mc}^{h*}=1} \sum_{o \in \mathcal{O} | x_{omc}^{h*}=1} (1 - x_{omc})\right) \tag{52}$$

**Theorem 1.** Cut (52) is a valid optimality cut.

**Proof.** In the sub-sequent iterations, the two following cases can happen:

**Case 1.** the same assignments are made to machine $m$, then the makespan will be bounded to $C_{max}^{hm*}$.

**Case 2.** At least one of $x_{omc}$ that $x_{omc}^{h*} = 1$ takes 0, then the right side of the cut will take a negative value, and it will be non-binding.

As a result, this cut will not eliminate any new feasible solution in the subsequent iterations, and in the case of a same assignment, it will limit the makespan to $C_{max}^{hm*}$.

#### 5.1.5. LBBD improvement

Based on several preliminary experiments on large-sized problems, it was found that the developed LBBD algorithm can be time-consuming, mostly due to the complexity of the MP and its large solution space. In this regard, an optimality gap of $\varepsilon$% can be set in the MP for its early termination. In this case, the optimality gap (between the best upper bound and lower bound) of the original model will be equal to or less than $\varepsilon$% [72]. Moreover, if the ratio of the number of orders to machines is large (e.g., greater than 40), setting a time limit can lead to a faster termination of SPs and more iterations in a limited computational time for the LBBD. However, it may hurt to reach an optimal solution by generating less accurate optimality cuts that remove some optimal solutions. Nevertheless, these improvement strategies are very effective in finding near-optimal solutions for large-sized instances in a limited computational time.

Presenting a suitable starting point through the warm start technique to MILP models can significantly enhance their performance by helping the solver to shrink the problem, and pruning some parts of the search tree. This starting point can be a complete or partial solution. As a result, a better solution may be obtained within a shorter computational time. This initial solution can be generated by another MILP model, a heuristic, a metaheuristic, or a hybrid method. In this paper, in each iteration, the solution obtained by the MP in the last iteration is used (assignment of orders to machine-configuration and batching of orders) as a starting point for the MP.

### 5.2. Warm-started MILP

This paper applies a two-step procedure to generate a partial initial solution to warm start the MILP. In the first step, the MP model presented in section 5.1.2 with a modification is solved to specify the assignment of orders to machine-configurations ($x_{omc}$), and utilized configurations ($w_{mc}$). This modification is about a better approximation of reconfiguration times in each machine. In the MP, the minimum reconfiguration time to a utilized configuration is calculated considering all configurations of the corresponding machine. While in Constraints (53)-(56), it is calculated according to utilized configurations on the machine. This leads to a tighter lower bound. Constraints (50)-(53) are not utilized in the MP due to the presence of big-M constraints that deteriorate the speed of solving procedure, and the MP needs to be solved several times. On the contrary, for warm starting the MILP model, it is solved just once. As a result, a tighter lower is useful to achieve a better initial solution.

$u_{mcc'}$ 1 if $\mathcal{T}_{mcc'}$ is the minimum reconfiguration time to configuration $c'$among the utilized configuration on machine $m$; 0, otherwise.

$$mrt_{mc'} \geq \mathcal{T}_{mcc'} - G \cdot (1 - u_{mcc'}), \qquad \forall m \in \mathcal{M}, \forall c, c' \in \mathcal{C}_m | c \neq c' \tag{53}$$

$$mrt_{mc'} \geq \mathcal{T}_{m0c'} - G \cdot (1 - u_{m0c'}), \qquad \forall m \in \mathcal{M}, \forall c' \in \mathcal{C}_m \tag{54}$$

$$\sum_{c \in \mathcal{C}_m \cup 0 | c \neq c'} u_{mcc'} = w_{mc'}, \qquad \forall m \in \mathcal{M}, \forall c' \in \mathcal{C}_m \tag{55}$$

$$\sum_{c' \in \mathcal{C}_m | c \neq c'} u_{mcc'} \leq |\mathcal{C}_m| \cdot w_{mc}, \qquad \forall m \in \mathcal{M}, \forall c \in \mathcal{C}_m \tag{56}$$

Constraints (53)–(56) calculate the minimum possible reconfiguration time to a utilized configuration ($mrt_{mc'}$) based on the relevant utilized and initial configurations of the corresponding machine. In these constraints, $u_{mcc'}$ is a binary variable equals to 1 if $\mathcal{T}_{mcc'}$ is the minimum reconfiguration time to configuration $c'$among the utilized configuration on machine $m$; 0, otherwise.

In the second step, for each machine the following MILP model is solved to determine the batching of the assigned orders on each machine ($x_{obm}$) with the objective of minimizing the number of batches (Equation (57)). Eventually, the assignment of orders to batch-machine-configuration combinations can be achieved ($x_{obmc}$). This partial solution is utilized as starting point to the MILP model.

$$Minimize \sum_{b \in \mathcal{B}} \sum_{c \in \mathcal{C}_m | w_{mc}=1} y_{bmc} \tag{57}$$

Subject to:

$$\sum_{b \in \mathcal{B}} x_{obm} = 1, \quad \forall o \in \mathcal{O} | \sum_{c \in \mathcal{C}_m} x_{omc} = 1 \tag{58}$$

$$\sum_{c \in \mathcal{C}_m | w_{mc}=1} y_{bmc} \leq 1, \quad \forall b \in \mathcal{B} \tag{59}$$

$$\sum_{o \in \mathcal{O} | x_{omc}=1} x_{obm} \leq |\mathcal{O}| \cdot y_{bmc}, \quad \forall b \in \mathcal{B}, \forall c \in \mathcal{C}_m | w_{mc} = 1, BL_{mc} = 0 \tag{60}$$

$$\sum_{o \in \mathcal{O} | x_{omc}=1} x_{obm} \leq y_{bmc}, \quad \forall b \in \mathcal{B}, \forall c \in \mathcal{C}_m | w_{mc} = 1, BL_{mc} = 1 \tag{61}$$

$$\sum_{o \in \mathcal{O} | \sum_{c \in \mathcal{C}_m} x_{omc}=1} x_{obm} \cdot OA_o \leq MPA_m, \quad \forall b \in \mathcal{B} \tag{62}$$

$$x_{obm}, y_{bmc}, z_{mcc'} \in \{0,1\}, \tag{63}$$

$$\forall o \in \mathcal{O} | \sum_{c \in \mathcal{C}_m} x_{omc} = 1, \forall b \in \mathcal{B}, \forall c, c' \in \mathcal{C}_m | w_{mc} = w_{mc'} = 1, c \neq c'$$

The descriptions of Constraints (58)-(62) are as the same as Constraints (41)-(45). Constraint (63) specifies the variables of the model.

## 6. Computational experiments

This section concentrates on providing comprehensive computational experiments. In this regard, the way of generating random instances is illustrated first. Afterward, the performance of the LBBD method in solving the random instances is contrasted with that of MILP, CP, and WMILP models. Finally, the sensitivity of results to the variation in the key parameters' values is assessed, which can provide valuable insights.

To run the developed solution methods in each experiment, the IBM CPLEX Optimization Studio 20.1.0 solver is employed, and its related DOcplex.MP and DOcplex.CP libraries are imported into the Python 3.8 programming language. All the related scripts are executed on a computer with an Intel (R) Core (TM) i5–7300HQ processor and 12 GB of RAM.

### 6.1. Instance generation

In order to evaluate and compare the performance of the developed solution methods, multiple instances are generated randomly, inspired by data presented in [42, 73]. These instances include different numbers of orders {50, 100, 200, 300, 400, 600, 800}, configurations per machine {5, 10, 15, 20}, and machines {5, 10, 20, 30 40}. Note that the efficiency of an algorithm may be measured as a function of the size of the instance, i.e., the number of orders. The experiments in our study are specifically designed to include large-scale industrial instances with up to 800 orders, 20 configurations, and 40 machines and satisfy the design features including being: comprehensive and discriminant when multiple algorithms are compared. The instances are signified with $|\mathcal{O}|$-$|\mathcal{C}|$-$|\mathcal{M}|$ to represent the number of orders, the number of specific configurations belonging to each machine, and the number of machines. For example, 50-5-5 represents an instance with 50 orders, 5 configurations per machine, and 5 machines. As explained in Section 1, by adding or removing modules to a machine (e.g., changing the head of a robotic arm) different configurations with different production capabilities can be attained. As a result, it is possible to process an order with multiple eligible machine-configuration combinations ($\mathcal{E}_{omc} = 1$) as it is detailed in Table 2. The manner of the random instance generation is described in Table 2.

### 6.2. Performance evaluation of the developed solution methods

20 medium- and large-sized instances are solved by the developed MILP, CP, and WMILP models, as well as the LBBD method, under 10-minute and 60-minute computational time limits. It should be noted that a gap of 1% in solving the master problem and a time limit of 1 minute in solving SPs were incorporated into the LBBD method. The results are presented in Table 3 and Table 4, including the best lower bound (Best_LB) and best objective (Best_Obj) found by the methods, Gap with the Best_LB, and Related Percentage Deviation (RPD) with the Best_Obj. Calculating the Gap and RPD is presented in Equations (64) and (65), respectively. Accordingly, the superiority of the LBBD over other methods is evident in both 10-minute and 60-minute time limits, excepting 100-5-5 in the 10-minute limit and 50-5-10 and 100-5-5 in the 60-minute limit.

$$Gap = \frac{Objective - Best\ lower\ bound}{Objective} \times 100 \tag{64}$$

$$RPD = \frac{Objective - Best\ Objective}{Best\ Objective} \times 100 \tag{65}$$

**Table 2.** The manner of the random instance generation

| **Parameters** | **Values** |
|---|---|
| $OA_o$ | $Discrete\ uniform\ distribution$ (75,200) |
| $MPA_m$ | $Ceil$ ($normal\ distribution$ (500,150)) |
| $OPT_{omc}$ | $Discrete\ uniform\ distribution$ (20,100) $\times \varphi_{mc}$ |
| $\varphi_{mc}$ | $machine - configuration\ speed\ coefficient$,<br>$Uniform\ distribution$ (0.8,1.2) |
| $BST_{mc}$ | $Discrete\ uniform\ distribution$ (6,8) |
| $BL_{mc}$ | $On\ each\ machine - configuration, probability$ $(BL_{mc} = 1) = 0.1$ |
| $\mathcal{E}_{omc}$ | $For\ each\ order, first\ selecting\ a\ random\ number\ of\ machine, then\ on\ each\ machine,$<br>$randomly, between$ 1 $and$ 75% $of\ its\ configurations\ are\ eligible$ $(\mathcal{E}_{omc} = 1)$ |
| $\mathcal{T}_{mcc'}$ | $Discrete\ uniform\ distribution$ (15,30) |

To better contrast the performance of the developed methods, the average Gap and RPD obtained by each method in the 10-minute and 60-minute time limits are depicted in Fig. 5 and Fig. 6, respectively. The LBBD, as the best method, managed to obtain solutions with a remarkable average Gap of 2.12 (in the 10-minute limit) and 1.88 (in the 60-minute limit). It can be inferred that the LBBD can reach near-optimal solutions in the early stages of its executions, and no major improvement occurs later. This is mostly due to the quality of SPs' makespan approximation in the MP model. In addition, the WMILP model also exhibits a decent performance by obtaining an 8.49% Gap (in 10 mins) and a 5.63% Gap (in 60 mins). Moreover, the MILP and CP models do not demonstrate a promising performance regarding the Gap. Furthermore, with respect to the average RPD shown in Fig. 6, the supremacy of the LBBD method is also distinct from the other methods. Furthermore, it is observed that the average RPD of the MILP, CP, and WMILP are improved during their execution.

Based on the results, it can be concluded that the developed LBBD is an efficient solution method for the problem under consideration since it can provide near-optimal solutions for medium- and large-sized problems within a reasonably short computational time.

**Table 3.** Comparing the performance of the developed MILP, CP, WMILP, and LBBD in the 10-minute time limit.

| No. | $\lvert\mathcal{O}\rvert$-$\lvert\mathcal{C}\rvert$-$\lvert\mathcal{M}\rvert$ | Best_LB | Best_Obj | MILP | | CP | | WMILP | | LBBD | |
|---|---|---|---|---|---|---|---|---|---|---|---|
| | | | | Gap | RPD | Gap | RPD | Gap | RPD | Gap | RPD |
| 1 | 50-5-5 | 598.61 | 603.59 | 2.27% | 1.48% | 14.38% | 15.83% | 1.26% | 0.44% | **0.83%** | **0.00%** |
| 2 | 50-10-5 | 584.04 | 600.34 | 9.05% | 6.96% | 11.59% | 10.04% | 5.92% | 3.41% | **2.72%** | **0.00%** |
| 3 | 50-5-10 | 285.17 | 302.12 | 9.86% | 4.71% | 13.78% | 9.47% | 7.19% | 1.70% | **5.61%** | **0.00%** |
| 4 | 50-10-10 | 312.85 | 326.54 | 10.68% | 7.26% | 20.27% | 20.17% | 7.49% | 3.56% | **4.19%** | **0.00%** |
| 5 | 100-5-5 | 1079.36 | 1093.57 | 3.75% | 2.55% | 9.85% | 9.48% | **1.30%** | **0.00%** | 1.46% | 0.16% |
| 6 | 100-10-5 | 1128.77 | 1142.74 | 13.28% | 13.90% | 20.44% | 24.16% | 3.18% | 2.02% | **1.22%** | **0.00%** |
| 7 | 100-5-10 | 591.04 | 609.55 | 7.65% | 4.99% | 18.22% | 18.57% | 8.64% | 6.13% | **3.04%** | **0.00%** |
| 8 | 100-10-10 | 601.27 | 616.75 | 6.99% | 4.81% | 26.59% | 32.80% | 9.43% | 7.64% | **2.51%** | **0.00%** |
| 9 | 200-5-5 | 2330.59 | 2357.83 | 11.03% | 11.10% | 12.83% | 13.39% | 1.23% | 0.08% | **1.16%** | **0.00%** |
| 10 | 200-10-5 | 2155.50 | 2177.08 | 11.55% | 11.94% | 26.49% | 34.69% | 4.03% | 3.17% | **0.99%** | **0.00%** |
| 11 | 200-5-10 | 1172.70 | 1197.93 | 14.10% | 13.96% | 20.51% | 23.15% | 5.84% | 3.97% | **2.11%** | **0.00%** |
| 12 | 200-10-10 | 1154.98 | 1173.57 | 22.24% | 26.56% | 32.91% | 46.70% | 12.51% | 12.49% | **1.58%** | **0.00%** |
| 13 | 300-5-10 | 1745.34 | 1772.20 | 12.52% | 12.57% | 18.43% | 20.73% | 5.14% | 3.82% | **1.52%** | **0.00%** |
| 14 | 300-10-10 | 1643.81 | 1662.91 | 14.59% | 15.74% | 35.12% | 52.36% | 17.31% | 19.54% | **1.15%** | **0.00%** |
| 15 | 300-5-20 | 866.80 | 883.50 | 19.64% | 22.09% | 29.03% | 38.23% | 15.36% | 15.91% | **1.89%** | **0.00%** |
| 16 | 300-10-20 | 868.70 | 897.72 | 34.33% | 47.36% | 45.23% | 76.69% | 17.39% | 17.14% | **3.23%** | **0.00%** |
| 17 | 400-5-10 | 2266.69 | 2297.63 | 34.77% | 51.25% | 28.85% | 38.66% | 1.64% | 0.30% | **1.35%** | **0.00%** |
| 18 | 400-10-10 | 2209.48 | 2241.26 | 25.11% | 31.63% | 33.86% | 49.05% | 11.68% | 11.62% | **1.42%** | **0.00%** |
| 19 | 400-5-20 | 1166.54 | 1188.48 | 21.42% | 24.91% | 34.21% | 49.19% | 15.64% | 16.36% | **1.85%** | **0.00%** |
| 20 | 400-10-20 | 1107.72 | 1136.86 | 25.79% | 31.30% | 49.22% | 91.88% | 17.68% | 18.36% | **2.56%** | **0.00%** |
| | **Average** | | | 15.53% | 17.35% | 25.09% | 33.76% | 8.49% | 7.38% | **2.12%** | **0.01%** |

**Table 4.** Comparing the performance of the developed MILP, CP, WMILP, and LBBD in the 60-minute time limit.

| No. | $\|\mathcal{O}\|$-$\|\mathcal{C}\|$-$\|\mathcal{M}\|$ | Best_LB | Best_Obj | MILP | | CP | | WMILP | | LBBD | |
|---|---|---|---|---|---|---|---|---|---|---|---|
| | | | | Gap | RPD | Gap | RPD | Gap | RPD | Gap | RPD |
| 1 | 50-5-5 | 598.61 | 603.59 | 2.20% | 1.40% | 12.18% | 12.93% | 1.26% | 0.44% | **0.83%** | **0.00%** |
| 2 | 50-10-5 | 584.04 | 598.14 | 5.93% | 3.80% | 10.92% | 9.61% | 5.77% | 3.62% | **2.36%** | **0.00%** |
| 3 | 50-5-10 | 285.17 | 298.08 | 9.40% | 5.59% | 10.30% | 6.65% | **4.33%** | **0.00%** | 5.45% | 1.19% |
| 4 | 50-10-10 | 312.85 | 322.55 | 12.45% | 10.79% | 19.43% | 20.39% | 6.48% | 3.72% | **3.01%** | **0.00%** |
| 5 | 100-5-5 | 1079.36 | 1090.92 | 2.45% | 1.43% | 6.84% | 6.20% | **1.06%** | **0.00%** | 1.28% | 0.23% |
| 6 | 100-10-5 | 1128.77 | 1142.74 | 7.30% | 6.55% | 14.06% | 14.94% | 2.86% | 1.69% | **1.22%** | **0.00%** |
| 7 | 100-5-10 | 591.05 | 609.55 | 7.36% | 4.67% | 11.39% | 9.43% | 5.16% | 2.24% | **3.04%** | **0.00%** |
| 8 | 100-10-10 | 601.28 | 615.37 | 7.88% | 6.07% | 21.62% | 24.67% | 7.04% | 5.11% | **2.29%** | **0.00%** |
| 9 | 200-5-5 | 2333.21 | 2357.26 | 3.44% | 2.51% | 8.62% | 8.31% | 1.12% | 0.10% | **1.02%** | **0.00%** |
| 10 | 200-10-5 | 2155.50 | 2174.75 | 11.52% | 12.02% | 14.92% | 16.49% | 2.26% | 1.41% | **0.89%** | **0.00%** |
| 11 | 200-5-10 | 1172.70 | 1193.26 | 4.22% | 2.60% | 15.30% | 16.03% | 3.98% | 2.35% | **1.72%** | **0.00%** |
| 12 | 200-10-10 | 1154.99 | 1171.89 | 24.82% | 31.10% | 25.35% | 32.02% | 9.73% | 9.18% | **1.44%** | **0.00%** |
| 13 | 300-5-10 | 1745.34 | 1769.27 | 11.76% | 11.80% | 16.89% | 18.70% | 4.53% | 3.32% | **1.35%** | **0.00%** |
| 14 | 300-10-10 | 1643.81 | 1661.94 | 16.25% | 18.10% | 28.02% | 37.41% | 8.39% | 7.97% | **1.09%** | **0.00%** |
| 15 | 300-5-20 | 866.80 | 883.40 | 22.13% | 26.01% | 25.13% | 31.05% | 9.43% | 8.34% | **1.88%** | **0.00%** |
| 16 | 300-10-20 | 869.56 | 894.18 | 21.17% | 23.36% | 33.13% | 45.43% | 13.07% | 11.86% | **2.75%** | **0.00%** |
| 17 | 400-5-10 | 2266.69 | 2286.49 | 9.71% | 9.79% | 18.05% | 20.97% | 1.40% | 0.54% | **0.87%** | **0.00%** |
| 18 | 400-10-10 | 2209.48 | 2235.31 | 24.49% | 30.91% | 28.20% | 37.66% | 11.27% | 11.41% | **1.16%** | **0.00%** |
| 19 | 400-5-20 | 1166.54 | 1188.48 | 20.81% | 23.94% | 23.23% | 27.86% | 11.11% | 10.42% | **1.85%** | **0.00%** |
| 20 | 400-10-20 | 1107.85 | 1131.05 | 16.27% | 16.98% | 35.56% | 52.01% | 2.40% | 0.35% | **2.05%** | **0.00%** |
| | **Average** | | | 12.12% | 12.47% | 19.00% | 22.44% | 5.63% | 4.20% | **1.88%** | **0.07%** |

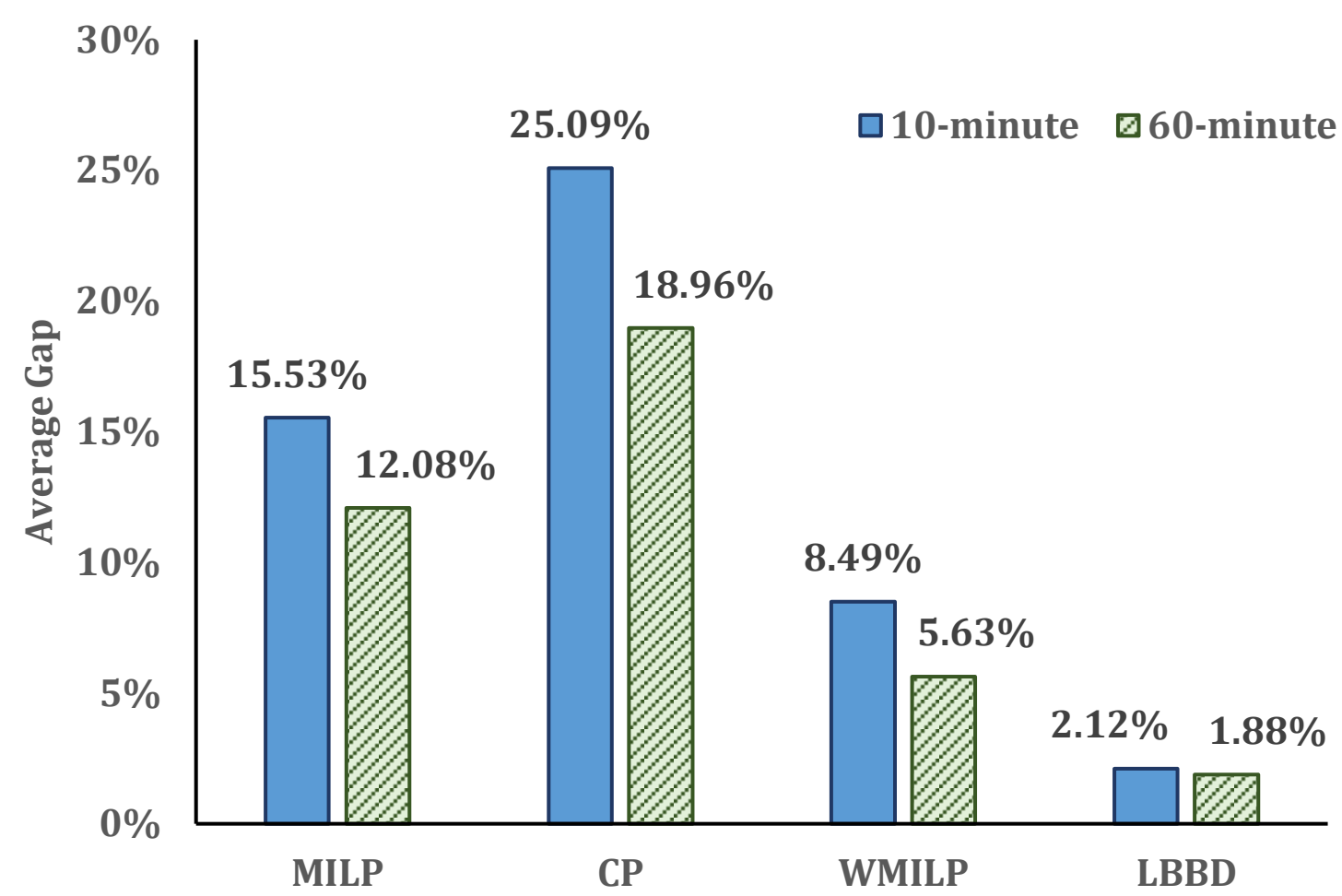

**Fig. 5.** Comparing the developed solution methods regarding the average Gap

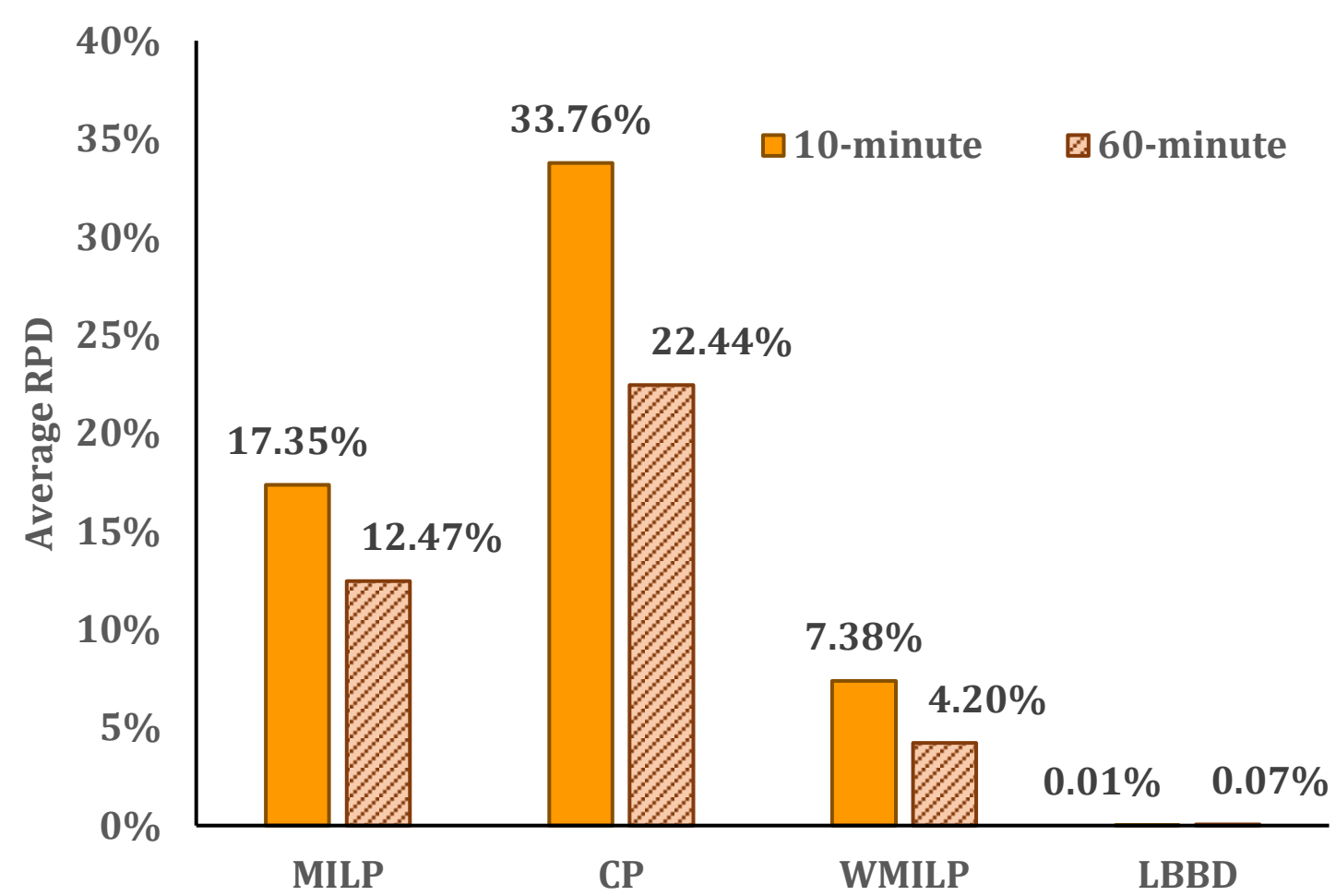

**Fig. 6.** Comparing the developed solution methods regarding the average RPD

Table 5 is also designed to test the computational limitations of the LBBD on large-sized instances. In Table 4, we observed that the number of orders impacts the solution quality. Therefore, we report the results for two extreme values of $|\mathcal{O}|$=600 and $|\mathcal{O}|$=800 as well as two values of 30 and 40 for $|\mathcal{M}|$. Specifically, in Table 5, we evaluate the performance of LBBD for four different values of $|\mathcal{C}|$ to analyze the complexity of reconfiguration decisions.

For large-sized industrial instances, the initial settings of the MP and SPs of LBBD do not always guarantee the best performance, as the MP may consume a large portion of the time limit to generate a solution. In such cases, the optimality gap of 1% used for solving the MP for its early termination in Table 4 is increased to 2% in Table 5 to better align with the LBBD method. This adjustment allows for faster termination of the MP and enables more iterations of the LBBD within the 60-minute time limit, although it may result in a lower-quality MP solution.

It is interesting to observe that the Gaps in Table 5 are relatively consistent with those in Table 4 for $|\mathcal{C}|$=5, indicating that the LBBD remains effective in large-sized instances where MILP, CP, and WMILP become computationally impractical. When $|\mathcal{C}|$ alternates between 10 and 20, the LBBD continues to perform effectively in finding near-optimal solutions for large-sized instances. Nevertheless, we can observe that gaps slightly diverge for larger values of $|\mathcal{C}|$ in Table 5. For example, let us analyze 600-20-30 and 800-20-40, respectively. For 600-20-30, setting a 2% optimality gap in the MP helps solve the problem effectively, achieving a solution Gap of 2.69%. This setting also leads to 28 iterations, which are sufficient. In contrast, the largest instance, with a size of 800-20-40, demonstrates that the LBBD may not always achieve a reasonable solution when the MP optimality gap is fixed at 2%. This is evident from the solution Gap of 4.33% for this instance in Table 5. More importantly, the number of iterations is only one, which is not sufficient. An approach to address this issue is to further increase the 2% optimality gap of the MP. However, a larger optimality gap in the MP risks transforming the LBBD from an exact method into a heuristic approach.

**Table 5.** Performance of the LBBD for large-sized instances with varying numbers of configurations, a termination condition of 2% optimality gap in the MP, and a 60-minute time limit for the LBBD.

| | | $\|\mathcal{C}\|$=5 | | $\|\mathcal{C}\|$=10 | | $\|\mathcal{C}\|$=15 | | $\|\mathcal{C}\|$=20 | | Average | |
|---|---|---|---|---|---|---|---|---|---|---|---|
| **No.** | $\|\mathcal{O}\|$-$\|\mathcal{M}\|$ | **Gap** | **Iteration** | **Gap** | **Iteration** | **Gap** | **Iteration** | **Gap** | **Iteration** | **Gap** | **Iteration** |
| 1 | 600-30 | 2.07% | 45 | 2.01% | 48 | 2.09% | 31 | 2.69% | 28 | 2.21% | 38 |
| 2 | 800-40 | 1.67% | 19 | 2.42% | 19 | 3.15% | 6 | 4.33% | 1 | 2.89% | 11.25 |
| | **Average** | 1.87% | 32 | 2.21% | 33.5 | 2.62% | 18.50 | 3.51% | 14.50 | | |

The Gantt chart for instance 200-5-10, obtained by the LBBD method within the 60-minute time limit, is depicted in Fig. 7. In this figure, the batches of a machine with the same configuration are presented by the same color. In addition, reconfiguration times and setup times are shown with black and grey colors, respectively.

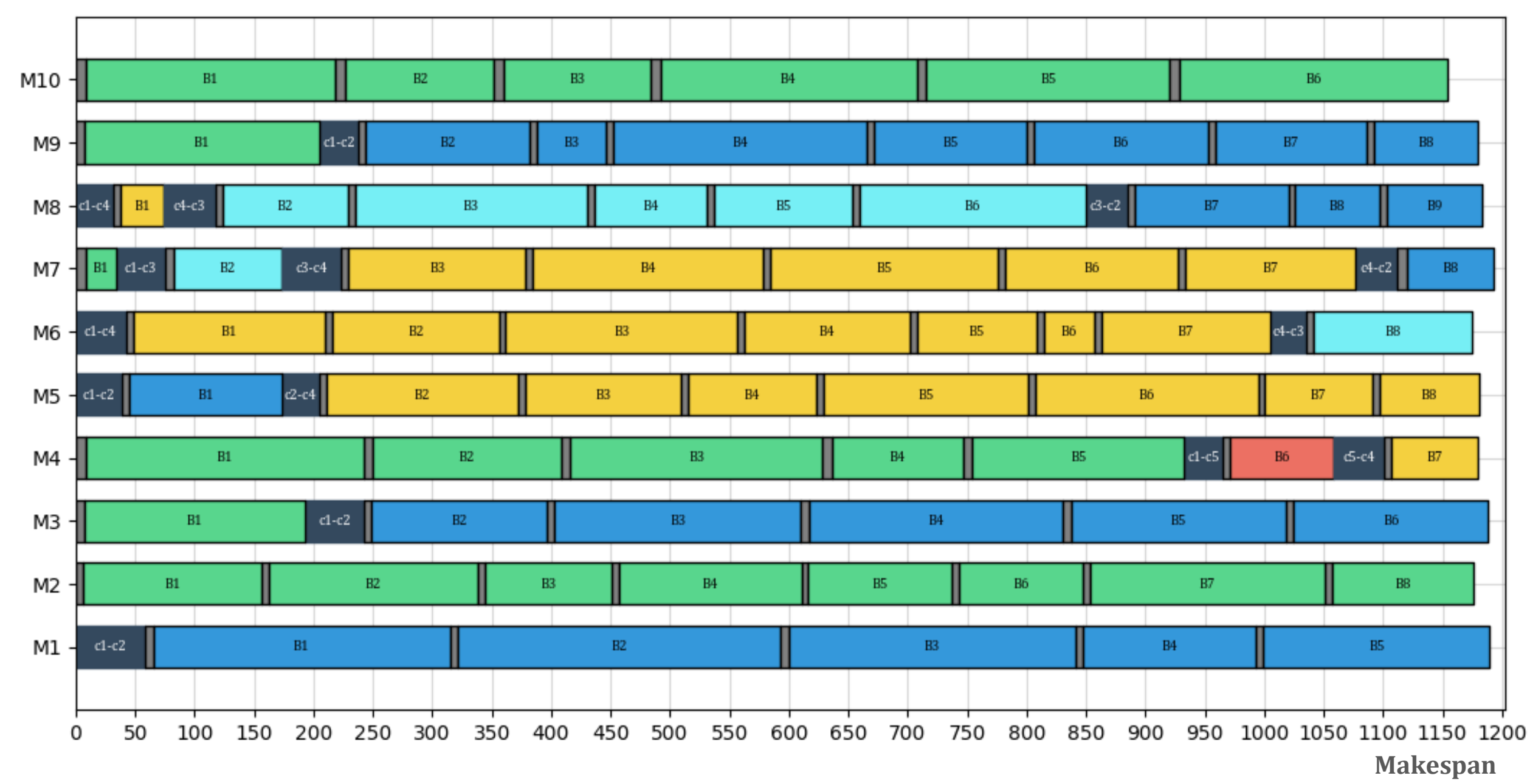

**Fig. 7.** Gantt chart for instance 200-5-10 obtained by the LBBD within 60-minute computational time

### 6.3. Sensitivity analysis

In this subsection, the impact of parameters' variation on the performance of the LBBD method and the scheduling outcome of the studied production system are explored. In this regard, the obtained solution gap of the LBBD method is regarded as its performance indicator. Furthermore, the achieved makespan is considered as the schedule performance indicator. To perform the analysis, the instance structure of 50-5-5 is considered. The analysis involves variations in the number of configurations per machine (3, 5, 7), number of machines (5, 7, 10), percentage of the eligible machine-configurations (30%, 50%, 70%), reconfiguration time (-50%, 0, 50%), setup time (-50%, 0, 50%), processing area (-50%, 0, +50%), and processing time variance (-50%, 0, +50%). For each experiment, 3 instances with different relevant parameters' values are generated and solved by the LBBD method, and the average results are reported. The analysis is illustrated in the following subsections.

#### *6.3.1. Impact of the parameters' variations on the Gap*

Fig. 8 demonstrates the performance of the LBBD method in terms of the variation of different parameters. As it can be perceived, increasing the number of configurations, number of machines, percentage of the eligible machine-configurations, reconfiguration time, setup time, and processing area, and decreasing the processing time variance result in higher Gaps. Accordingly, variations in the number of machines (Fig. 8(b)), reconfiguration time (Fig. 8(d)), and setup time (Fig. 8(e)) have the largest impacts on the obtained solution Gaps. For instance, increasing the number of machines from 5 to 10 leads to about a 6% surge in the Gap (Fig. 8(b)). Moreover, the changes in the value of the processing area (Fig. 8(f)) and processing time variance (Fig. 8(g)) have the least effect on the LBBD performance.

Overall, the LBBD method exhibits a rather robust performance against parameters' variations.

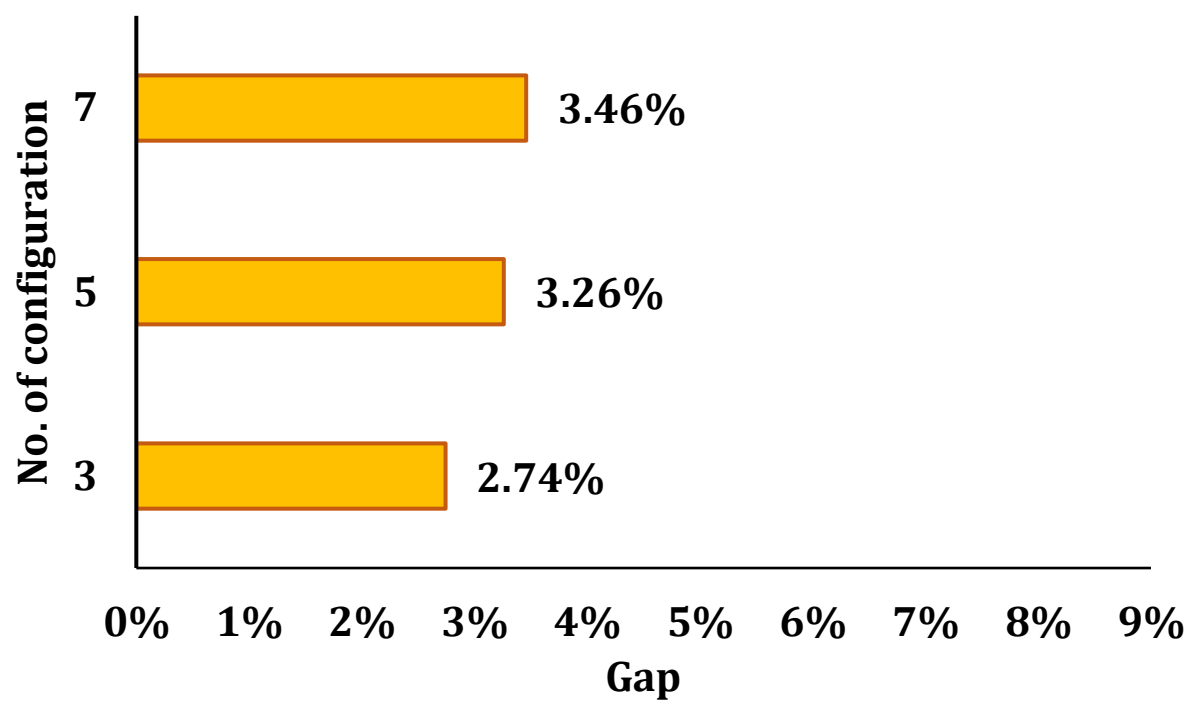

Fig. 8(a)

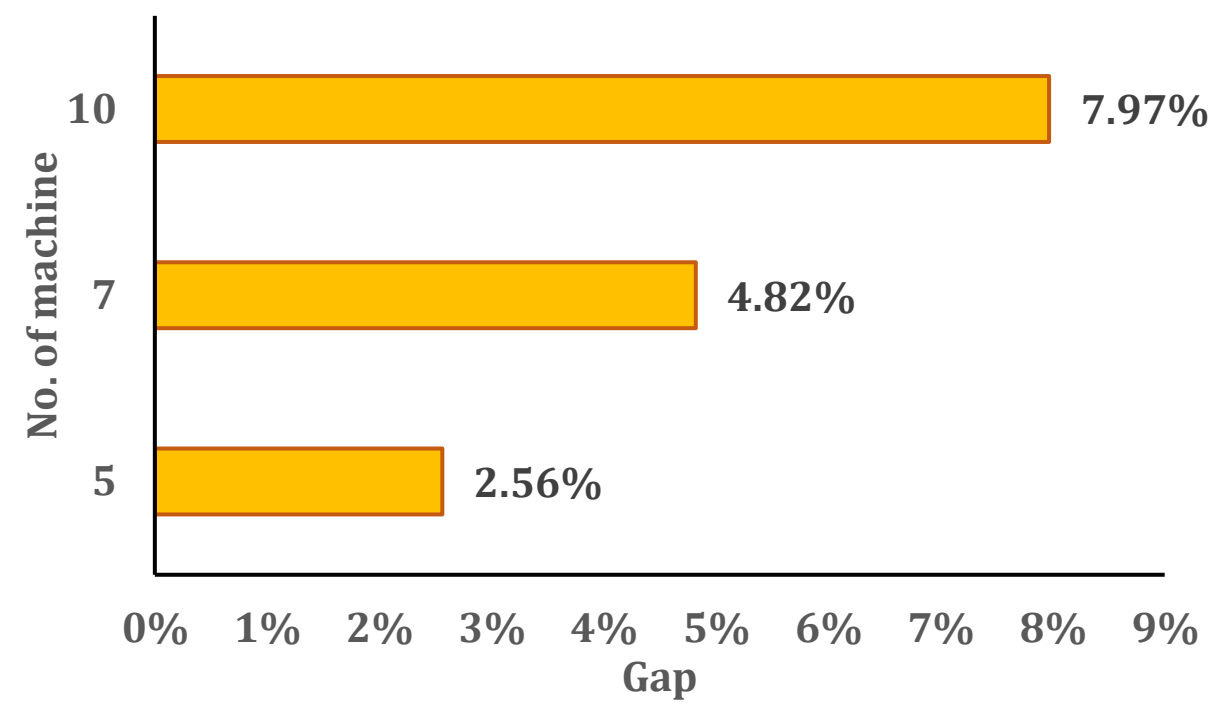

Fig. 8(b)

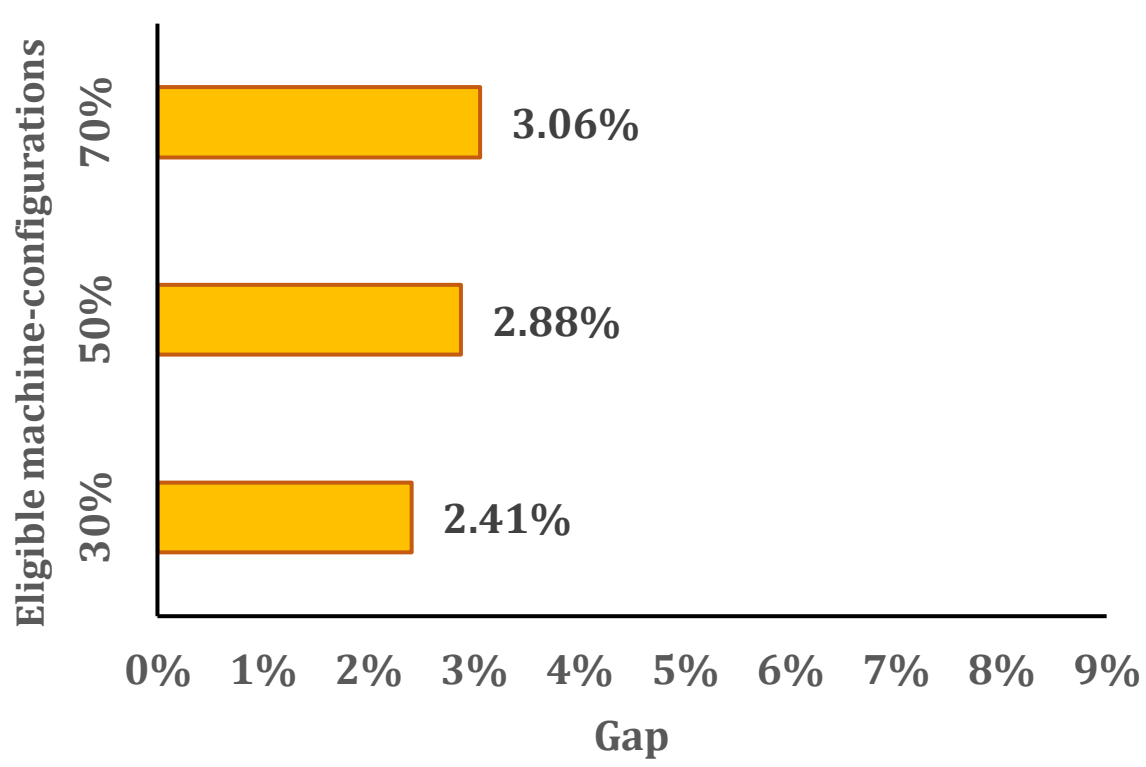

Fig. 8(c)

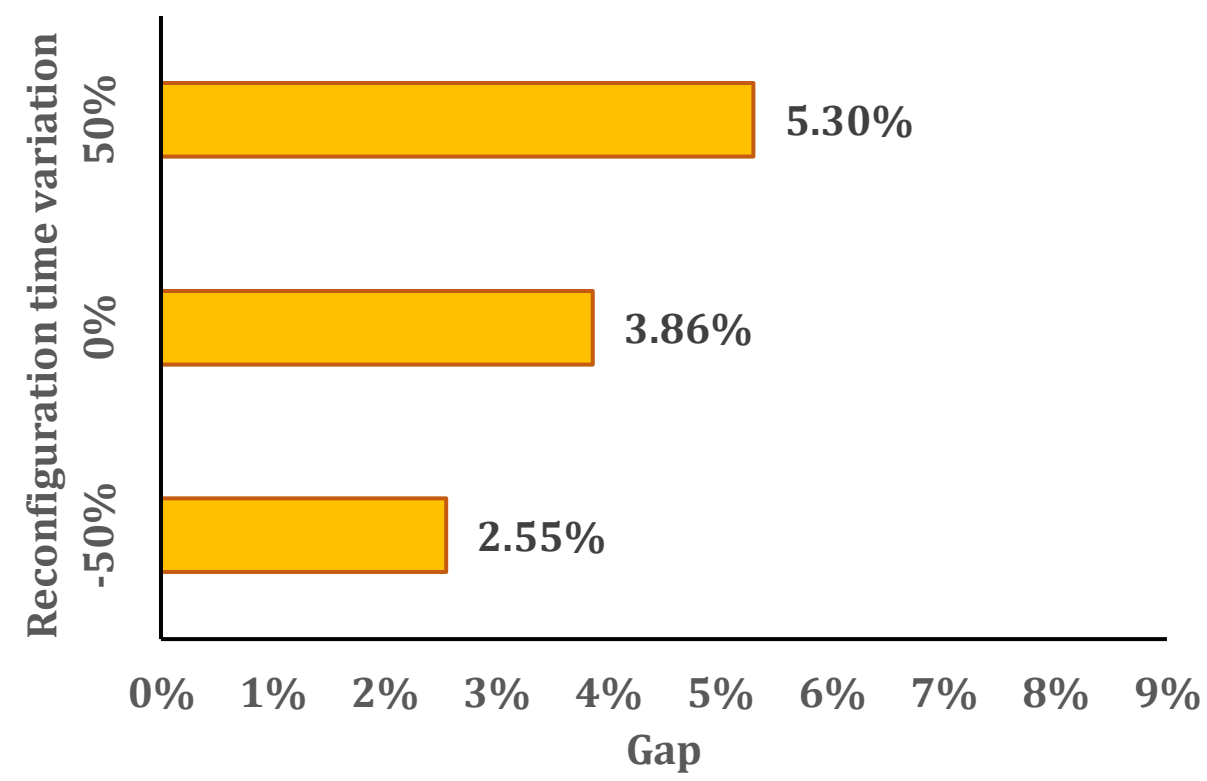

Fig. 8(d)

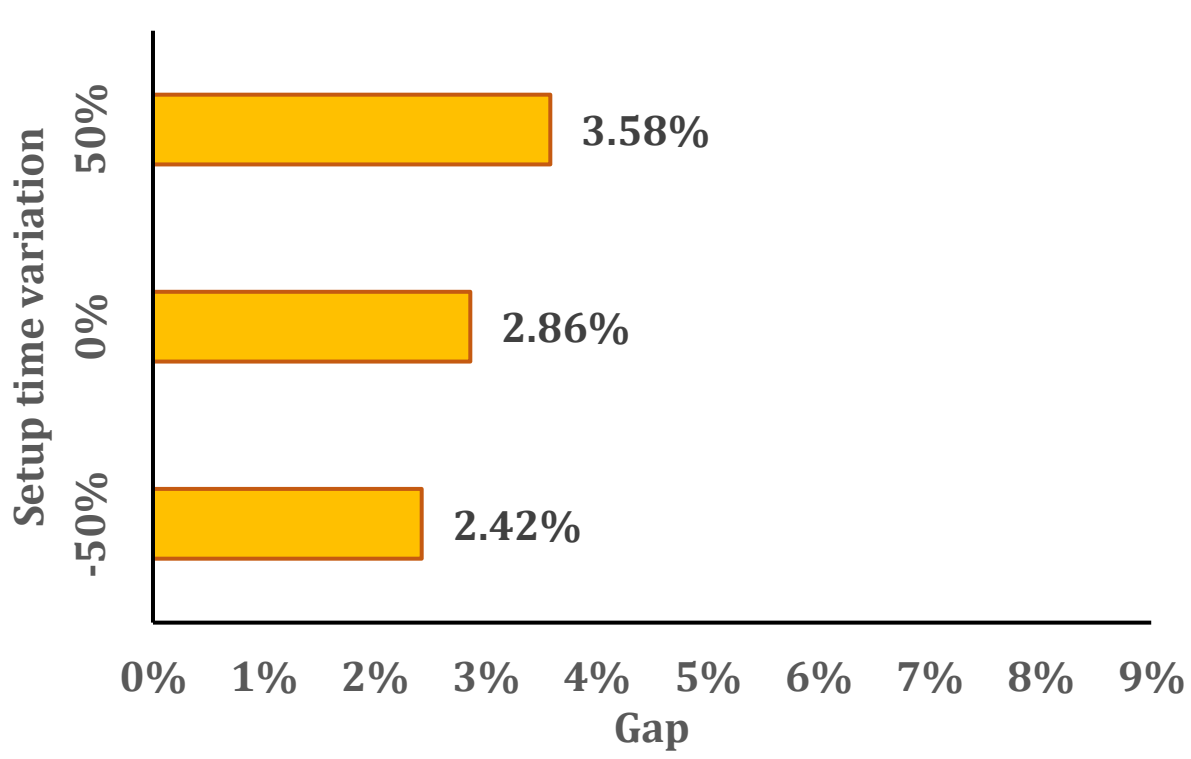

Fig. 8(e)

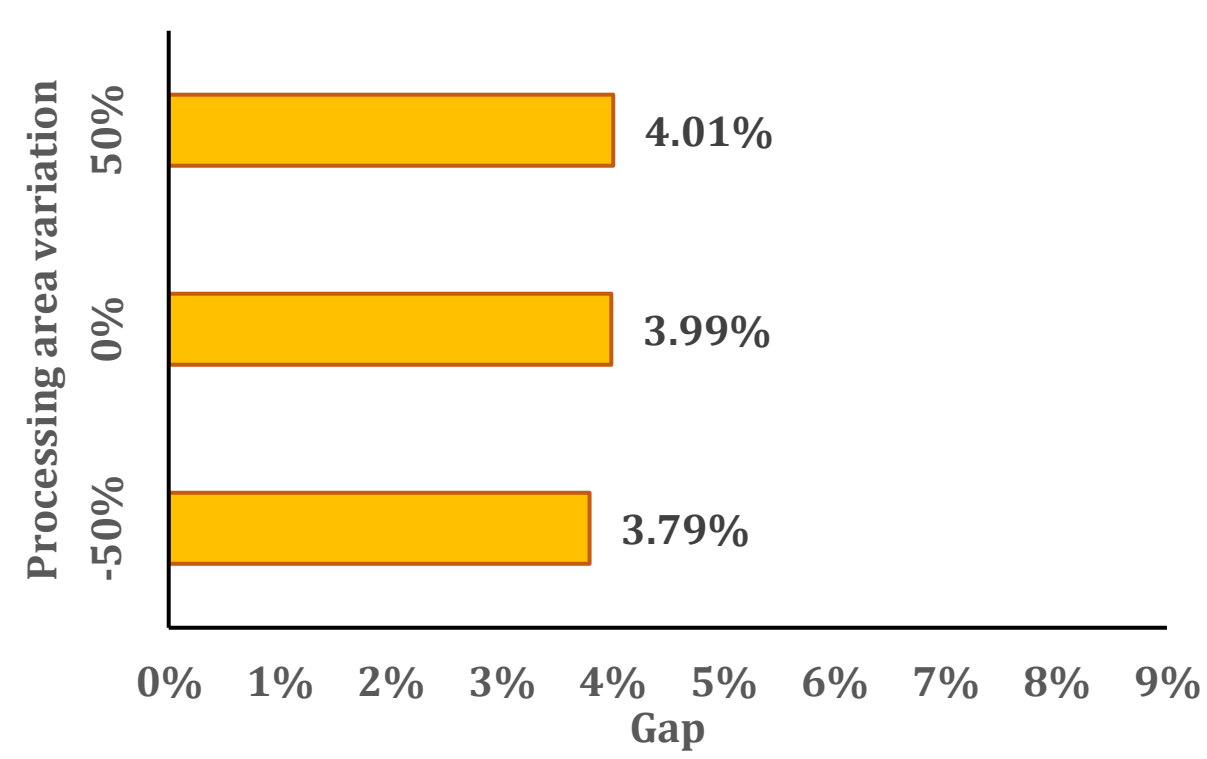

Fig. 8(f)

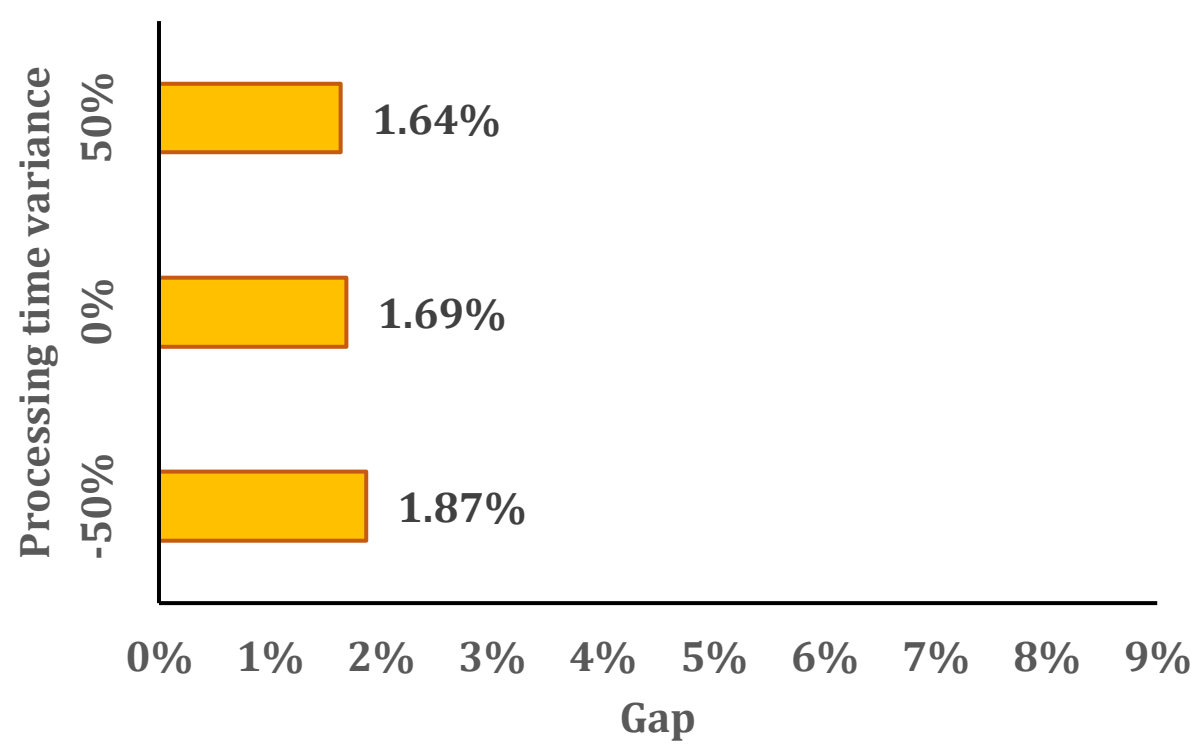

Fig. 8(g)

**Fig. 8.** Impact of the parameters' variations on the Gap

### *6.3.2. Impact of the parameter variation on the makespan*

Fig. 9 concerns the sensitivity analysis based on the efficiency of schedules. In this regard, Fig. 9(a) shows that the number of configurations has no major impact on the makespan, whereas changing the number of machines from 5 to 10 sharply reduces the makespan by about 47% (Fig. 9(b)). In contrast to Fig. 9(a), Fig. 9(c) also shows that the increase in the percentage of eligible machine-configuration combinations from 30% to 70% leads to about a 3% reduction in the makespan. Thus, employing more versatile machines' modules enhances the flexibility and efficiency of the HMRSs, which should be pondered considering technical and budget limits.

According to Fig. 9(d) and Fig. 9(e), alternations in the reconfiguration time and setup time have a meaningful effect on the efficiency of the HMRSs. For instance, a 50% reduction in reconfiguration time leads to about a 5.6% improvement in the makespan.

Fig. 9(f) reveals that the increase in the machine's processing area can strengthen the HMRSs' efficiency due to the reduction in the total setup times of machines. For instance, about 2% improvement can be achieved by a 50% increase in the processing area of machines.

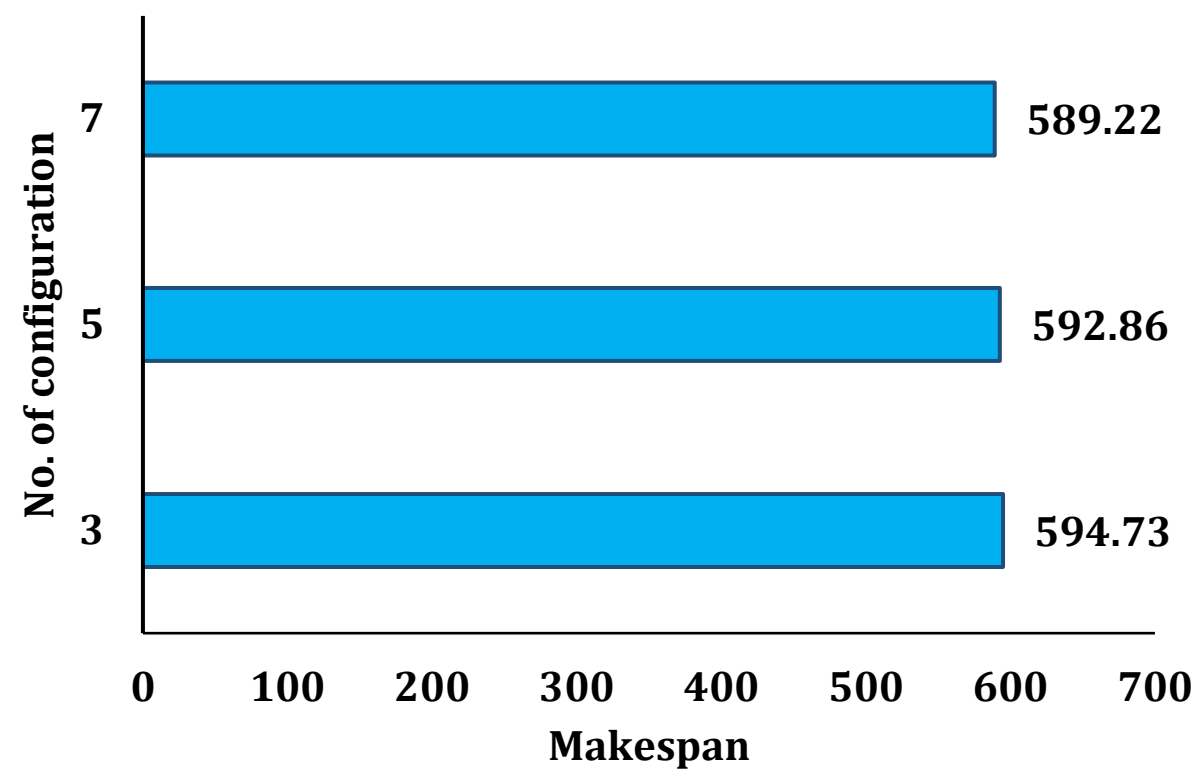

Fig. 9(a)

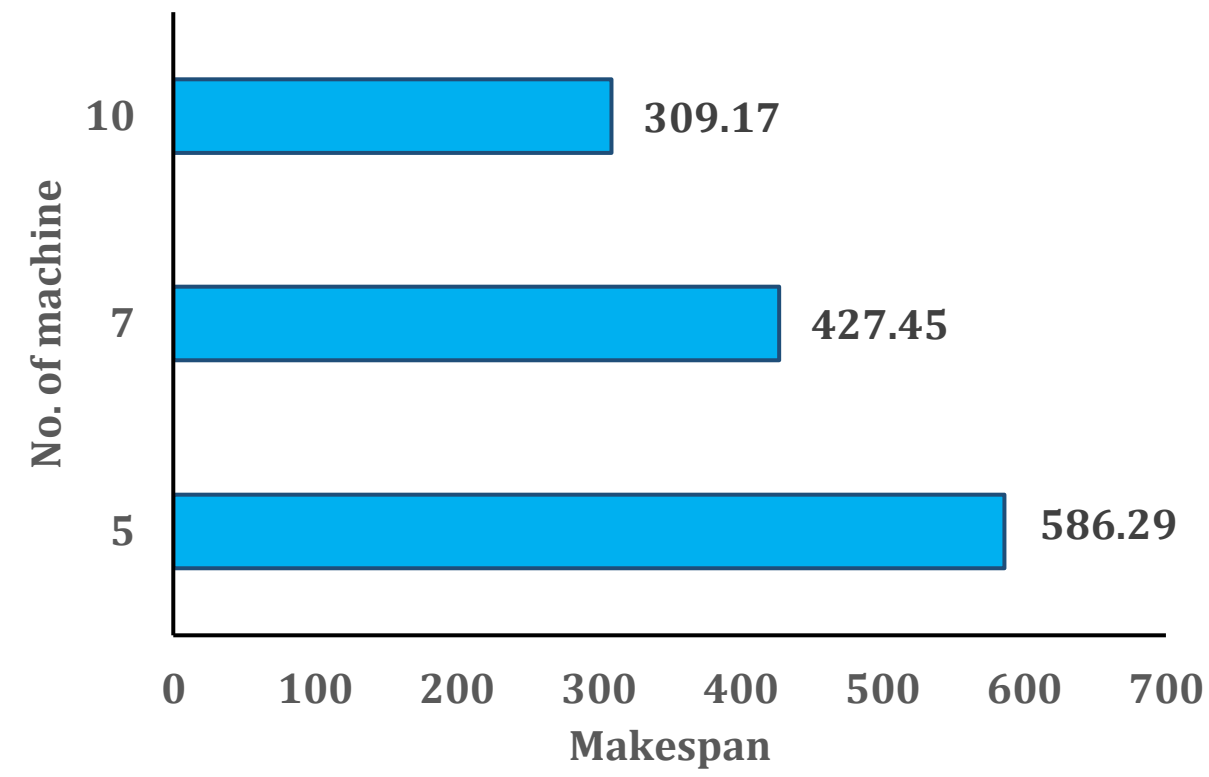

Fig. 9(b)

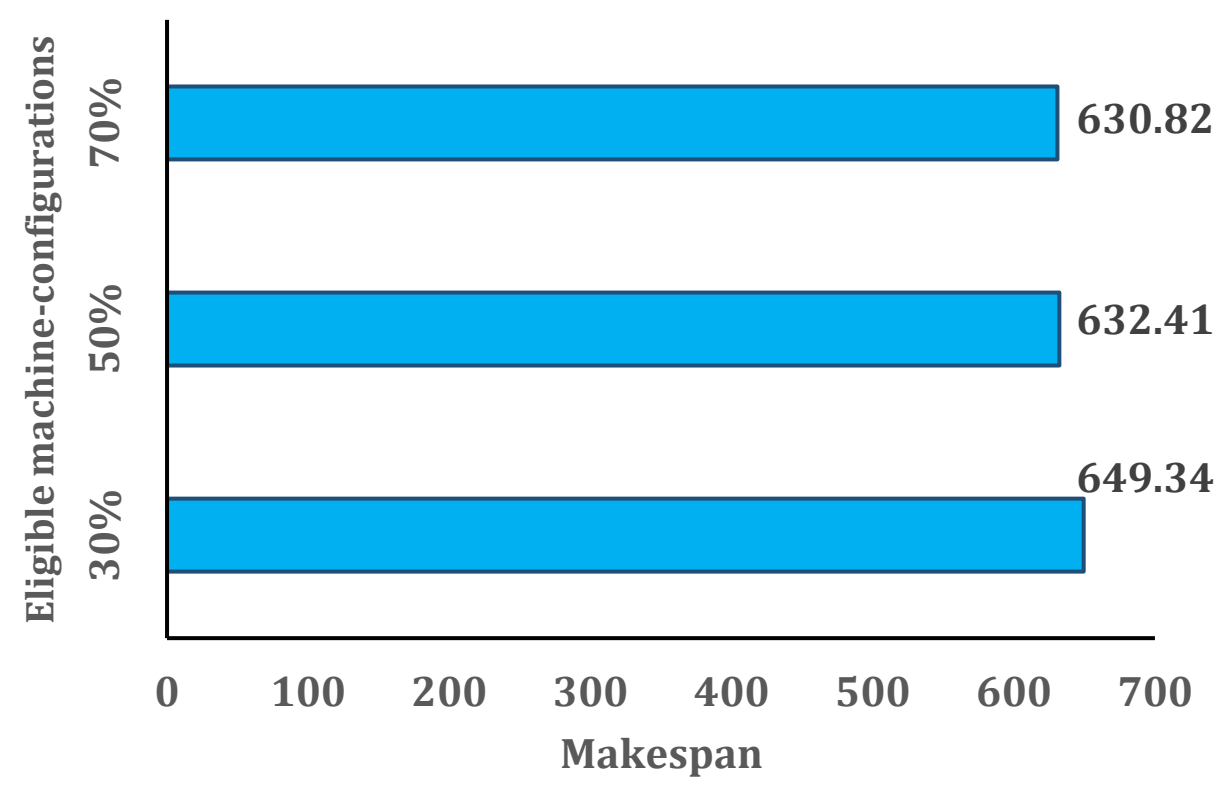

Fig. 9(c)

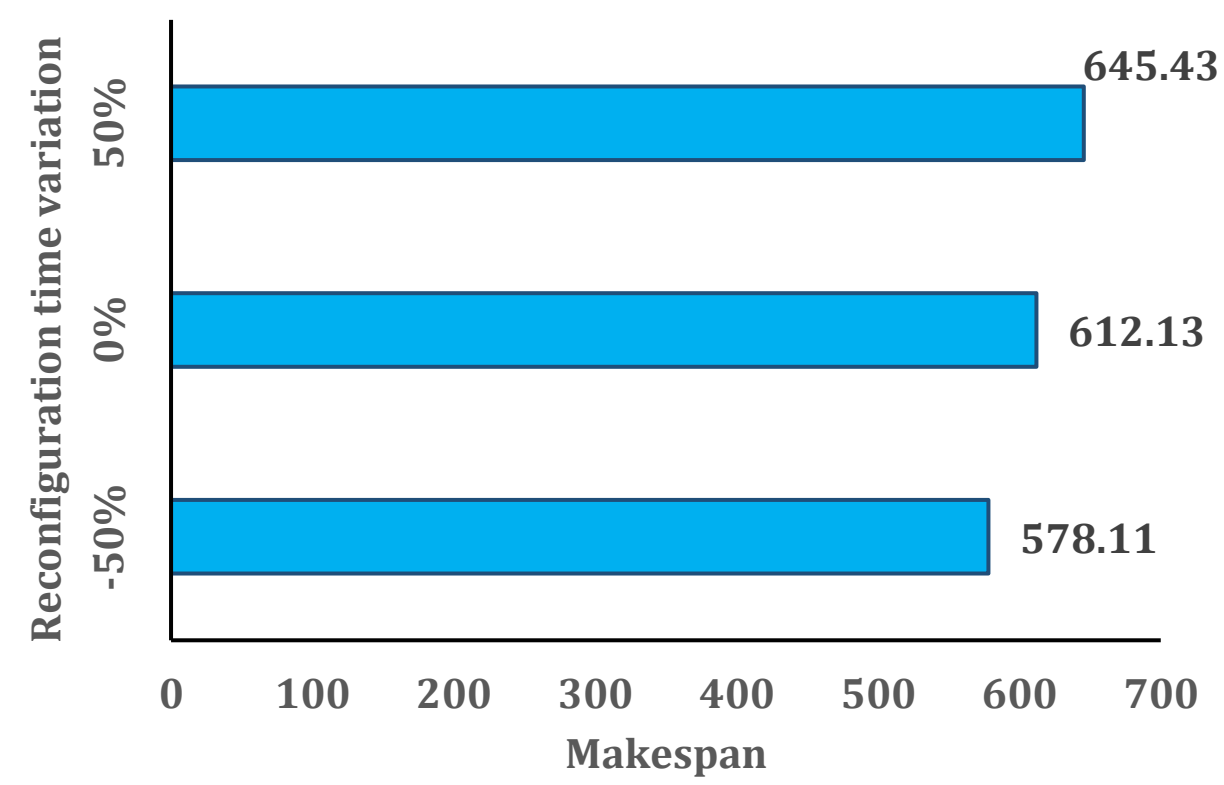

Fig. 9(d)

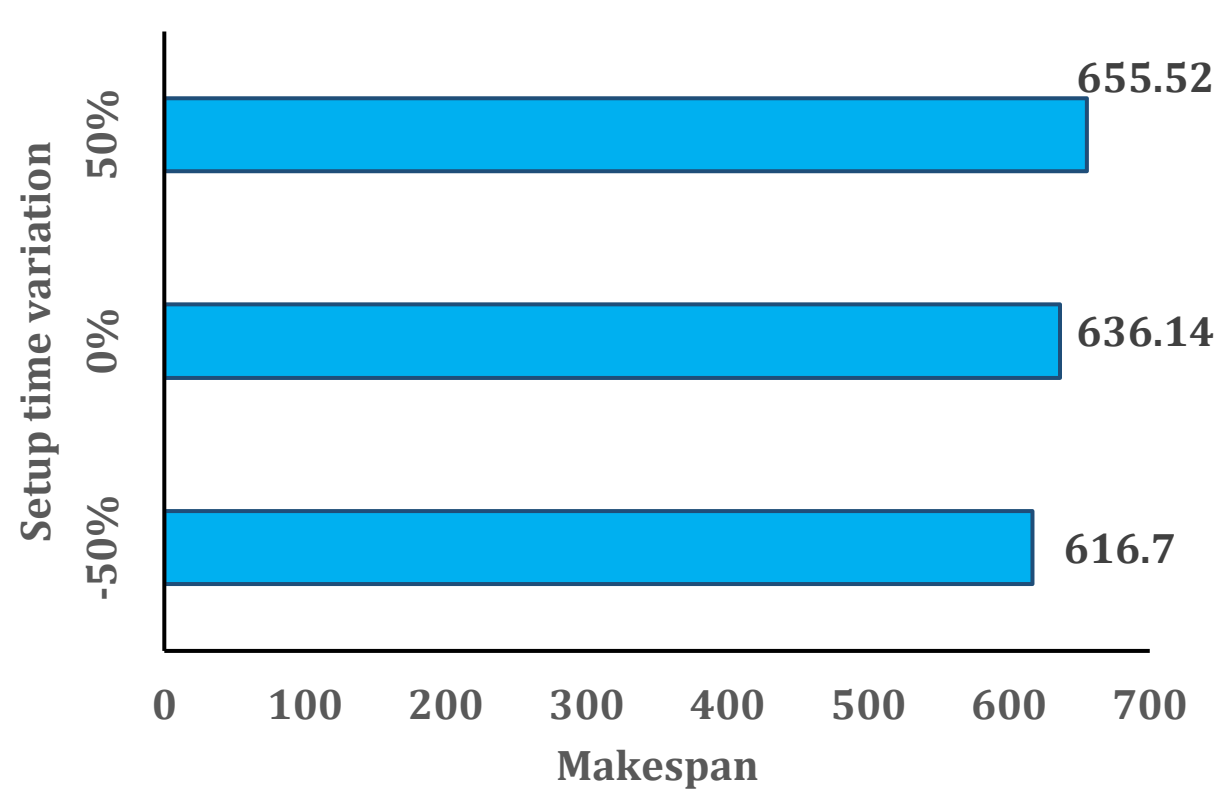

Fig. 9(e)

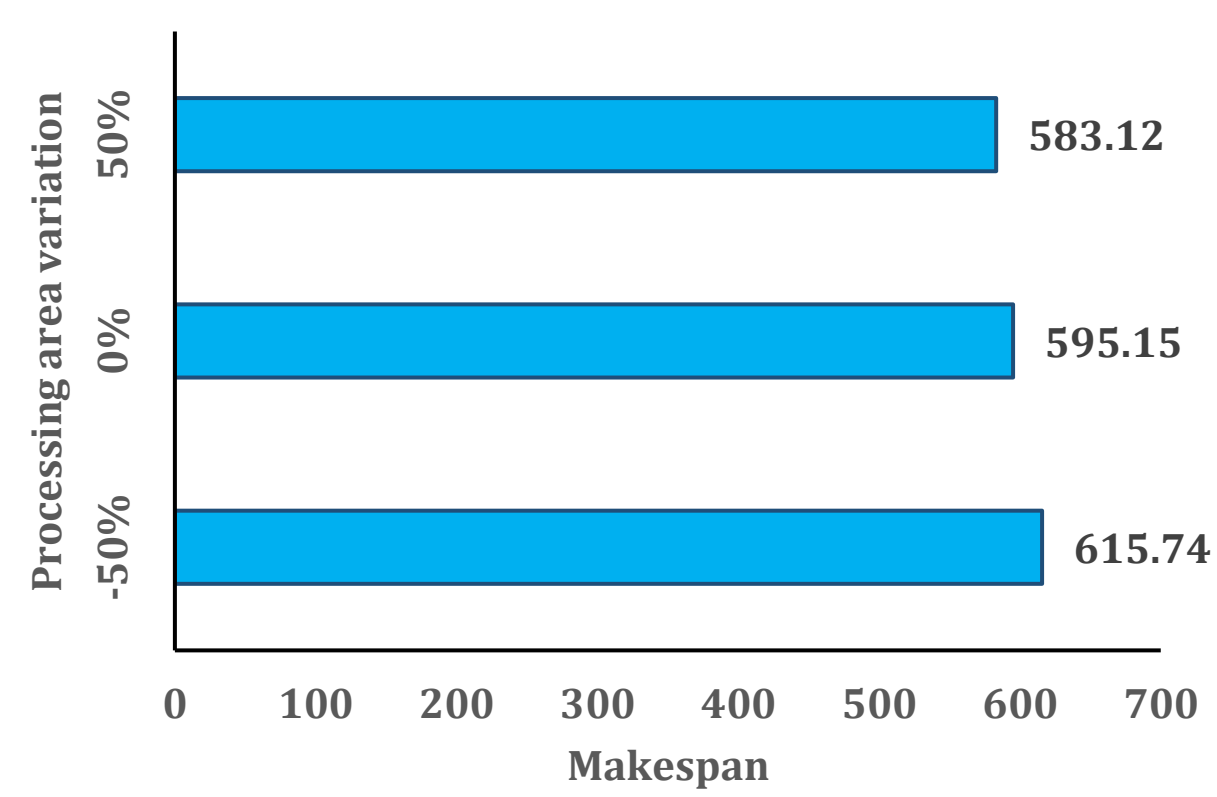

Fig. 9(f)

**Fig. 9.** Impact of the parameters' variations on the makespan

*6.4. Managerial insights and discussions*

To this end, we shed light on the effect of parameters' variation on the performance of the LBBD method, and the quality of the schedule by analyzing the solution Gap and makespan, respectively. Let us now consider Fig. 8 and Fig. 9 simultaneously in order to obtain additional managerial insights into the results presented in this section.

Proceeding in this manner reveals two types of parameters: low and high-priority parameters. The former type of parameters requires less attention, as they have no major influence on the Gap and makespan concurrently, i.e., the number of configurations, eligible machine-configuration combinations, processing area variation, and processing time variance. The conclusion from this comparison is that technical and budget allocation constraints should not be prioritized based on these types of parameters.

Conversely, the latter types of parameters should receive special attention from decision-makers as they can be related to strategic and tactical planning. For example, the relationship between the Gap and makespan performance indicators in Fig. 8(b) and Fig. 9(b) is strongly conflicting when the number of machines is changed. Hence, we can conclude that the policy-maker should decide on the right number of machines such that an appropriate balance is achieved between the Gap and makespan. Deciding on the right number of machines is crucial in strategic and tactical decisions to balance underutilization and responsiveness.

Fig. 8(d) and Fig. 9(d) reveal that no conflict exists between the Gap and makespan indicators when the analysis is limited to the reconfiguration time variation. Therefore, providing appropriate modules for machines that can be easily and effortlessly reconfigured is among managerial decisions that should be made only based on technical and budget constraints. Likewise, Fig. 8(e) and Fig. 9(e) establish a conflict-free environment in the HMRS, posing no difficulty regarding the setup time variation. Setup is among the managerial decisions that should be made meticulously based on technical and budget constraints. Furthermore, providing workers with proper training programs can be promising in reducing both reconfiguration time and setup time, particularly in complex environments like HMRSs.

The results presented in this section align with those in Section 6.2, reinforcing the reliability of the findings. More precisely, the insights gained from Sections 6.2 and 6.3 are also applicable to various HMRSs for handling dynamic and real-time changes in scheduling parameters. The benefit of fast convergence of solutions (i.e., a gap of 2.12% with the Best_LB within the 10-minute limit, as shown in Table 3 and Fig. 5) for dynamic scheduling of HMRSs is significant, as it directly impacts the HMRSs' ability to adapt to changing conditions and make timely decisions. It can be concluded that rapid convergence minimizes the time required to generate and implement new schedules in real-time and respond quickly to dynamic conditions, such as new orders or changes in demand.

## 7. Conclusions and future research

This study investigated a batch production scheduling problem in a hybrid manufacturing-remanufacturing system comprising parallel reconfigurable machines. Accordingly, novel MILP and CP models were devised for the problem considering the makespan as the performance measure. The problem is highly complex due to the simultaneous decisions regarding assigning orders to machine-configuration combinations, batching orders in machines, and scheduling orders. In this regard, an efficient logic-based benders decomposition method was developed. In addition, the presented MILP model was enhanced by the warm start technique.

The computational experiments revealed the superiority of the LBBD method over the other developed solution methods by achieving the average Gap and RPD of 1.88% and 0.07%, respectively. Moreover, the results expressed that the variations in the number of machines, reconfiguration time, and setup time had the largest impacts on the LBBD method's performance. Furthermore, concerning the sensitivity analysis based on the makespan, it can be concluded that the number of machines, percentage of eligible machine-configuration combinations, reconfiguration time, setup time, and processing area were significantly impactful. Thus, proper decision-making at strategic and tactical levels regarding the number, specifications, and versatility of machines and their modules respecting the technical and budget limits, as well as suitable worker training are of great importance in the HMRSs.

Human well-being and energy consumption are among the most challenging issues confronting nowadays manufacturing industry. In this line, incorporating human-related objectives like ergonomics, workload balancing, and worker satisfaction can be explored to extend this study. In addition, considering different energy consumptions due to different configurations and the presence of time of use energy cost in the problem are promising directions for future studies.

## Declaration of Competing Interest

The authors declare that they have no known competing financial interests or personal relationships that could have appeared to influence the work reported in this paper.

## Acknowledgment

This work was co-funded by the European Union under the project ROBOPROX (reg. no. CZ.02.01.01/00/22_008/0004590).